\documentclass[preprint,noshowpacs,noshowkeys,floatfix,aps,pra]{revtex4-2}
\usepackage{graphicx}
\usepackage[dvipsnames]{xcolor}
\usepackage{amssymb}
\usepackage{amsmath}
\usepackage{stmaryrd}
\usepackage{tikz}
\usetikzlibrary{shapes.geometric,shapes.misc,positioning,calc}
\definecolor{Blue}  {RGB}{  0, 100, 200}
\definecolor{Green} {RGB}{150, 200,   0}
\definecolor{Orange}{RGB}{250, 150,   0}
\usepackage{caption}
\usepackage{subcaption}

\newtheorem{definition}{Definition}

\newtheorem{remark}{Remark}
\newtheorem{example}{Example}

\begin{document}

\title{Solving the time-independent Schr\"{o}dinger equation for chains of coupled excitons and phonons using tensor trains}

\author{Patrick Gel\ss{}}
\affiliation{
Institut f\"{u}r Mathematik, Freie Universit\"{a}t Berlin \\ Arnimallee 3--9, D-14195 Berlin, Germany}

\author{Rupert Klein}
\affiliation{
Institut f\"{u}r Mathematik, Freie Universit\"{a}t Berlin \\ Arnimallee 3--9, D-14195 Berlin, Germany}

\author{Sebastian Matera}
\affiliation{
Institut f\"{u}r Mathematik, Freie Universit\"{a}t Berlin \\ Arnimallee 3--9, D-14195 Berlin, Germany}

\author{Burkhard Schmidt}
\email{burkhard.schmidt@fu-berlin.de}
\affiliation{
Institut f\"{u}r Mathematik, Freie Universit\"{a}t Berlin \\ Arnimallee 3--9, D-14195 Berlin, Germany}

\date{\today}

\begin{abstract}

We demonstrate how to apply the tensor-train format to solve the time-independent Schr\"{o}dinger equation for quasi one-dimensional excitonic chain systems with and without periodic boundary conditions.
The coupled excitons and phonons are modeled by Fr\"{o}hlich-Holstein type Hamiltonians with on-site and nearest-neighbor interactions only.
We reduce the memory consumption as well as the computational costs significantly by employing efficient decompositions to construct low rank tensor-train representations, thus mitigating the curse of dimensionality.
In order to compute also higher quantum states, we introduce an approach which directly incorporates the Wielandt deflation technique into the alternating linear scheme for the solution of eigenproblems.
Besides systems with coupled excitons and phonons, we also investigate uncoupled problems for which (semi-)analytical results exist.
There, we find that in case of homogeneous systems the tensor-train ranks of state vectors only marginally depend on the chain length which results in a linear growth of the storage consumption. 
However, the CPU time increases slightly faster with the chain length than the storage consumption because the alternating linear scheme adopted in our work requires more iterations to achieve convergence for longer chains and a given rank. 
Finally, we demonstrate that the tensor-train approach to the quantum treatment of coupled excitons and phonons makes it possible to directly tackle the phenomenon of mutual self-trapping.
We are able to confirm the main results of the Davydov theory, i.e., the dependence of the wavepacket width and the corresponding stabilization energy on the exciton-phonon coupling strength, though only for a certain range of that parameter. 
In future work, our approach will allow calculations also beyond the validity regime of that theory and/or beyond the restrictions of the Fr\"{o}hlich-Holstein type Hamiltonians.
\end{abstract}

\maketitle

\section{Introduction}
\label{sec:introduction}
Excitons are electron-hole pairs held together by Coulomb interaction which can be generated in semiconducting materials by interaction with light of suitable wavelength. 
In the simplest case, electrons are excited from the (highest) occupied to the (lowest) unoccupied molecular orbital, or from valence to conduction band. 
Regarded as quasi-particles in solid-state materials, excitons can transport energy without transporting net electric charge.
The efficiency of excitonic energy transfer (EET) plays a crucial role in natural and artificial light harvesting systems. 
For example, in photovoltaic solar cells, excitons have to migrate from "antenna" sites of  efficient light absorption to active interfaces such as electrodes or embedded catalytic sites in order for charge separation to occur. 
The present work focuses on organic semiconductors (OSCs) materials such as $\pi$--conjugated small molecules or polymer chains~\cite{Bardeen2014,Mikhnenko2015,Hedley2016,Zhugayevych2015,Kranz2016,Nelson2020}.
For a variety of opto-electronic applications, OSC-based materials offer many potential technological advantages over anorganic materials, such as mechanical flexibility, easier and cheaper fabrication, and less adverse environmental impacts.
Due to the lower dielectricity, excitons in OSCs are typically more localized than in anorganic materials, i.e., they are closer to the limiting case of Frenkel excitons than Wannier--Mott excitons.

An important feature of the photo-induced exciton dynamics in organic semiconductors is a strong exciton-phonon coupling (EPC)~\cite{Zhugayevych2015}.
Hence, for understanding the structure and dynamics of excitons, a quantitative investigation of their coupling to lattice vibrations (phonons) is of paramount importance.
Typically, EPC processes in solid OSC materials are modeled in terms of Fr\"{o}hlich-Holstein type Hamiltonians originally developed to describe electrons in a polarizable lattice~\cite{Devreese2009}.
These models, or minor modifications thereof, are also known as Huang-Rhys or Peierls models~\cite{Schroter2015}.
Moreover, there is a close analogy to the Davydov model describing the transport of amide I vibrational energy along $\alpha$-helical proteins~\cite{Scott1992,Georgiev2019}. 
We also note the similarity of the above model Hamiltonians to molecular Jahn-Teller Hamiltonians used in the theory of "vibronic" (i.e., vibrational-electronic) coupling where they are frequently used to model non-adiabatic quantum dynamics at conical intersections~\cite{Domcke2004}.

The structural and dynamic properties of excitons are quite sensitive to the physical parameters of the Fr\"{o}hlich-Holstein type Hamiltonians. 
In the simplest possible case of a one-dimensional solid with two electronic states state and one (acoustic) phonon mode per site and zero temperature, the parameter space is essentially spanned by two parameters, namely the coupling between neighboring excitons and the EPC parameters~\cite{Devreese2009,Zhugayevych2015,Sakkinen2015}.
This allows for a classification in terms of the relative strengths of the interactions where weak and strong couplings can be determined spectroscopically from the ratio of excitonic and vibrational bandwidths~\cite{Simpson1957}.
Even though it is hardly possible to mention all the investigations of the parameter or size dependence of  properties of Fr\"{o}hlich-Holstein type models, the main feature upon increasing the EPC strength is a cross-over from almost free electrons or excitons to self-trapped quasi-particles known as polarons or solitons, respectively~\cite{Scholes2010,Binder2018}.
This phenomenon of self-localization or "phonon dressing" appears universally: 
Not only electrons or excitons can be trapped through interaction with lattice phonons~\cite{Gerlach1991,Devreese2009} but also amide I vibrational energy quanta can become localized through interaction with the stretching modes of the hydrogen bonds that stabilize the $\alpha$-helices in proteins~\cite{Scott1992,Georgiev2019}.

Even though the Fr\"{o}hlich-Holstein type Hamiltonians represent only the simplest models for EPC in OSC materials, and even after decades of intense research, quantum-mechanical calculations for these systems are far from being trivial.
With the corresponding Schr\"{o}dinger equations not being exactly solvable, one has to resort to numerical procedures.
Such approaches, however, become quite challenging once the number of interacting sites is increased~\cite{Schroter2015}.
This is because quantum states, as well as Hamiltonians, can be considered as tensors, i.e., multidimensional generalizations of vectors and matrices. 
The number of elements of those tensors -- and, thus, their storage consumption -- grows exponentially with the number of dimensions. 
This so-called \emph{curse of dimensionality} applies also to analyzing complex quantum systems, i.e., solving the time-independent Schr\"{o}dinger equation (TISE).
There, the computational effort grows exponentially so that systems of increasing dimensionality quickly become intractable when applying conventional grid methods.
An alternative to solving the TISE are quantum Monte Carlo methods \cite{Mishchenko2000,Titantah2001}. 
However, their use is typically limited to the calculation of very low-lying states.
For the solution of the time-dependent Schr\"{o}dinger equations (TDSE), the multiconfiguration time-dependent Hartree (MCTDH) method and its multilayer (ML) extensions are known to be highly efficient algorithms for propagating wavepackets in high dimensions~\cite{Beck2000,Wang2003,Meyer2009}.
Indeed, ML-MCTDH has been applied successfully in simulations of the excitonic energy transfer in OSC polymer chains~\cite{Binder2018,DiMaiolo2020}.

In the present work, we want to exploit low-rank tensor decompositions of quantum state vectors and Hamiltonian matrices to mitigate the \emph{curse of dimensionality}.
Over the last decades, the interest in tensor decompositions has been growing constantly since proposed decomposition techniques have shown that it is possible to tackle high-dimensional systems in many application areas, e.g., quantum physics~\cite{White1992, Meyer2009,Beck2000}, chemical reaction dynamics~\cite{Kazeev2014, Gelss2016}, and machine learning~\cite{Gelss2019, Klus2019}. 
That is, different tensor formats such as the \emph{canonical format}~\cite{Hitchcock1927}, the \emph{Tucker format}~\cite{Tucker1964}, and the \emph{tensor-train format}~\cite{Oseledets2009a} allow simulation and analysis of high-dimensional problems without an exponential scaling of the memory consumption and/or computational effort, provided they satisfy certain structural constraints.

Tensor formats are certainly not a cure of the curse of dimensionality for high-dimensional problems in general. However, the restriction to nearest neighbor (NN) interactions of coupled excitons and phonons in the model quantum Hamiltonians for quasi one-dimensional systems investigated here suggests the use of the tensor-train (TT) format, also known as matrix product state representation.
Indeed, recently developed techniques based on TT decompositions can provide a robust and
 efficient numerical framework for the solution of the TDSE~\cite{Borrelli2016,Borrelli2017,Greene2017} and the Liouville-von Neumann equation~\cite{Borrelli2019}.
Application of TT techniques to solve the TISE are scarce and -- to the best of our knowledge -- limited to calculations of the ground state~\cite{Orus2014}.
The TT format ensures not only a non-exponential dependence on the number of dimensions but also the numerically stable computation of solutions of best approximation problems~\cite{Holtz2012b}. 
The idea of the TT format is to decompose a high-dimensional tensor into a network of low-dimensional tensors -- called \emph{TT cores} -- which are coupled by the \emph{TT ranks}. 
In~\cite{Gelss2017}, we derived a systematic TT decomposition for systems comprising only NN interactions which shall be the basis for the present work. 
One of the main advantages of this so-called SLIM decomposition is that the TT ranks of homogeneous systems do not depend on the number of sites of the network, resulting in a linear scaling of the computational complexity with the number of sites.

Typically, the applications of the TT format require the approximation of solutions of eigenvalue problems or systems of linear equations, i.e., in our case the solutions of the TISE or TDSE, respectively. 
A standard method for solving such problems in the TT format is the \emph{alternating linear scheme} (ALS)~\cite{Holtz2012}. 
The basic idea is to fix all components of the tensor network except for one. 
This yields a series of low-dimensional eigenvalue problems (or systems of linear equations), which can then be solved using conventional numerical methods. 
That is, ALS updates the cores iteratively while sweeping bidirectionally through the TT network.
However, ALS computes only eigentensors corresponding to the smallest/largest magnitude eigenvalue~\cite{Holtz2012} of a given symmetric TT operator $H$. 
There exist approaches for the approximation of the eigentensors corresponding to the $K$ smallest/largest eigenvalues such as the application of \emph{block TT format}~\cite{Dolgov2014} or the \emph{locally optimal block preconditioned conjugate gradient} (LOBPCG) method~\cite{Rakhuba2019}.
Here, we will follow a different route: In order to compute quantum states corresponding to the lowest energy levels, we introduce a modification of ALS based on an integrated \emph{Wielandt deflation}~\cite{Saad2011} which enables us to displace previously computed eigenvalues while keeping all other eigenvalues unchanged.
To avoid the explosion of the computational costs for higher excited states, which would arise in a straight-forward application of the Wielandt deflation, the manipulation of the Hamiltonian is directly incorporated into the ALS.

In the present work these techniques shall be employed to obtain solutions of the TISE describing the structure of the coupled excitons and phonons in quasi one--dimensional OSC systems. In a follow-up publication we will explore the use of TT decomposition and ALS techniques to also solve the TDSE used for modeling excitonic energy transfer dynamics.

\section{Model Hamiltonians}
\label{sec:model}
Here, we introduce simple model Hamiltonians for coupled excitons and phonons.
We restrict ourselves to the consideration of effectively one-dimensional, linear chains comprising $N$  exciton-supporting sites assuming NN interactions only.
In the context of photovoltaics applications, such models are suitable for the description of two classes of systems~\cite{Mikhnenko2015,Hedley2016}.
First, in conjugated polymer chains without major kinks or turns the excitonic sites interact with each other mainly via a Dexter mechanism ("through bond").
Second, in crystals of polycyclic aromatic molecules the coupling of excitons occurs mainly in the direction perpendicular to the planes of the stacked molecules, governed by the F{\"o}rster mechanism ("through space").

In the present work, we model the EPC employing Fr\"{o}hlich-Holstein type Hamiltonians originally developed for the interaction of electrons with a polarizable lattice \cite{Devreese2009}.
These models, or minor modifications thereof, also known as Fr{\"o}hlich, Huang-Rhys, or Peierls models, represent the simplest Hamiltonians accounting for electron-phonon coupling in organic semiconductors \cite{Zhugayevych2015}.
In passing, we also note the close analogy to the Davydov models describing the interaction of the amide I vibrations and the hydrogen bonds that stabilize the $\alpha$-helix of proteins~\cite{Scott1992,Georgiev2019}. 

To begin with, the total Hamiltonians can be written as
\begin{equation}
	H = H^{\mathrm{(ex)}} \otimes \mathbb{I}^{\mathrm{(ph)}} 
	+ \mathbb{I}^{\mathrm{(ex)}} \otimes H^{\mathrm{(ph)}} 
	+ H^{\mathrm{(ex-ph)}} 
	\label{eq:H_total}
\end{equation}
where the superscripts (ex) and (ph) stand for excitons and phonons and where $\mathbb{I}$ are identity operators on the respective particle spaces.
A simple one-exciton Hamiltonian for the interaction of local excitons along a linear or cyclic (chain or ring) system of $N$ (not necessarily identical) sites can be written in terms of (bosonic) exciton raising, $b_i^\dagger$, and lowering, $b_i$, operators
\begin{equation}
	H^{\mathrm{(ex)}} = \sum_{i=1}^N \alpha_i
	                    b_i^\dagger b_i
	                  + \sum_{i=1}^{N} \beta_i
										  \left(
											b_i^\dagger b_{i+1} + 
											b_i b_{i+1}^\dagger
											\right)
	\label{eq:H_ex}
\end{equation}
with local ("on site") excitation energy $\alpha_i$ for site $i$.
The nearest-neighbor (NN) coupling energy $\beta_i$ between site $i$ and $i+1$, often referred to as transfer or hopping integral, determines the exciton delocalization and mobility.
Within the framework of Kasha's theory for molecular crystals, a negative or positive sign of $\beta$ correlates with a head-to-tail (J--aggregate) or a side-by-side (H--aggregate) alignment of the constituting molecules, respectively~\cite{Hestand2017}.

Here and throughout the following, the last summand ($i=N$) of the NN coupling term (with indices $i+1$ replaced by 1) is used for cyclic systems with periodic boundary conditions only and is deleted otherwise.
Here, we are using Wannier (position space) representation, where excitons and vibrational degrees of freedom are labeled by site indices, rather than the Bloch (momentum space) representation, 
because of the finite number of sites usually encountered in organic systems~\cite{Bardeen2014}.
Moreover, the former one is more convenient for representation of NN interactions and for the use of tensor trains, see Sec.~\ref{sec:numerics}.

The Hamiltonian for the vibrational (phononic) degrees of freedom of a one-dimensional lattice can be written in terms of masses $m_i$, displacements $R_i$, and conjugated momenta $P_i$ in harmonic approximation
\begin{equation}
	H^{\mathrm{(ph)}} = \frac{1}{2}            \sum_{i=1}^N \frac{P_i^2}{m_i}
										+ \frac{1}{2} \sum_{i=1}^N m_i \nu_i^2        R_i^2
	                  + \frac{1}{2} \sum_{i=1}^N \mu_i \omega_i^2 \left( R_i-R_{i+1} \right)^2
	\label{eq:H_ph0}
\end{equation}
where each site $i$ is restrained to oscillate around its equilibrium position by a harmonic potential with frequency parameter $\nu_i$ and where the NN interaction between sites $i$ and $i+1$ is modeled by a harmonic oscillator with frequency parameter $\omega_i$ and corresponding reduced mass $\mu_i=m_im_{i+1}/(m_i+m_{i+1})$.
Here, it is assumed that the frequencies $\nu_i$ and $\omega_i$ are independent of the excitonic state of the system.
Moreover, atomic units with $\hbar=1$ are used throughout this work.

In analogy to the treatment of the excitons, we introduce second quantization for the phononic Hamiltonian of Eq.~(\ref{eq:H_ph0}) yielding
\begin{equation}
	H^{\mathrm{(ph)}} = \sum_{i=1}^N \tilde{\nu}_i
	                    \left(c_i^\dagger c_i + \frac{1}{2}\right)
	                  - \sum_{i=1}^N \tilde{\omega}_i
										\left( c_i^\dagger+c_i \right)
										\left( c_{i+1}^\dagger+c_{i+1} \right)
	\label{eq:H_ph2}
\end{equation}
where raising and lowering operators of (local) vibrations of site $i$ are indicated by $c_i^\dagger$ and $c_i$, respectively, and where we found that
\begin{eqnarray}
\tilde{\nu}_i&=&\sqrt{\nu_i^2+\frac{m_{i-1}}{m_i+m_{i-1}}\omega_{i-1}^2+\frac{m_{i+1}}{m_i+m_{i+1}}\omega_{i}^2} \nonumber \\
\tilde{\omega}_i&=&\frac{\mu_i\omega_{i}^2}{2\sqrt{m_i \tilde{\nu}_i m_{i+1} \tilde{\nu}_{i+1}}}
\label{eq:omg_E}
\end{eqnarray}
are the effective frequencies of single site and NN pair vibrations.

Linear models for the coupling of excitons and phonons can be formulated in terms of a Fr\"{o}hlich-Holstein type coupling Hamiltonian \cite{Devreese2009}
\begin{equation}
	H_\chi^{\mathrm{(ex-ph)}} 
	                  = \sum_{i=1}^N \chi_i
	                    b_i^\dagger b_i \otimes R_i
									 = \sum_{i=1}^N \tilde{\chi}_i
	                    b_i^\dagger b_i \otimes 
											\left( c_i^\dagger+c_i \right)
	\label{eq:H_couple_chi}
\end{equation}
where the coupling constant $\chi_i=\mathrm{d}\alpha_i / \mathrm{d}R_i$ gives the linearized $R_i$-dependence of exciton energy $\alpha_i$.
Here and in the following, a tilde notation is used for converted EPC constants, e.g., $\tilde{\chi}_i=\chi_i/\sqrt{2m_i\tilde{\nu}_i}$, bearing the dimension of an energy.
It is physically more intuitive, however, to express the coupling in terms of NN distances 
\begin{equation}
	H_\rho^{\mathrm{(ex-ph)}} = \sum_{i=1}^N \tilde{\rho}_i
	                    b_i^\dagger b_i \otimes \left[
											\left( c_{i+1}^\dagger+c_{i+1} \right) -
											\left( c_i^\dagger+c_i \right)
											\right] \quad .
	\label{eq:H_couple_rho}
\end{equation}
That expression gives a reasonable description, e.g., of excitons in protein helices where an amide I vibration couples much more strongly to the directly adjacent H-bond \cite{Scott1991}.
For semiconductor materials, however, one typically assumes equal coupling to the NN distances on both sides, giving rise to a symmetrized \textit{ansatz} for the EPC
\begin{equation}
	H_\sigma^{\mathrm{(ex-ph)}} = \sum_{i=1}^N \tilde{\sigma}_i
	                    b_i^\dagger b_i \otimes \left[
											\left( c_{i+1}^\dagger+c_{i+1} \right) -
											\left( c_{i-1}^\dagger+c_{i-1} \right)
											\right]
	\label{eq:H_couple_sig}
\end{equation}
which is suitable in case of a mirror symmetry of the individual sites \cite{Scott1992,Georgiev2019}.

Finally, we also consider Holstein-Peierls models with an additional term
\begin{equation}
	H_\tau^{\mathrm{(ex-ph)}} = \sum_{i=1}^N \tilde{\tau}_i
	                    \left(b_i^\dagger b_{i+1} +
											b_i b_{i+1}^\dagger \right) \otimes \left[
											\left( c_{i+1}^\dagger+c_{i+1} \right) -
											\left( c_i^\dagger+c_i \right)
											\right]
	\label{eq:H_couple_tau}
\end{equation}
where $\tau_i$ gives the linearized distance-dependence of exciton coupling energy $\beta_i$ introduced in Eq.~\eqref{eq:H_ex}.
These (non-local) couplings are related to the Peierls-instabilities potentially giving rise, e.g., to spontaneous dimerization of sites in a chain or ring. 
In passing, we note the analogy with molecular vibronic Hamiltonians modeling Jahn-Teller effects and conical intersections \cite{Domcke2004}.

\section{Tensor Train Decompositions}
\label{sec:numerics}
\subsection{Statement of the problem}

Herein, we deal with the question of how to obtain stationary quantum states for the Fr\"{o}hlich-Holstein type Hamiltonians introduced in the previous section.
Except for the case of dimers \cite{Sakkinen2015}, the corresponding time-independent Schr{\"o}dinger equation (TISE) is known to be not analytically solvable.
Hence, we have to resort to numerical techniques and we will, in particular, work with low-rank tensor decomposition techniques employing the tensor train format.
In passing, we note that similar approaches can also be applied to solve the time-dependent Schr{\"o}dinger equation (TDSE) which will be the subject of a forthcoming publication.

First, let us consider the underlying Hilbert spaces and their dimensions.
A single site excitonic state vector is represented as
\begin{equation}
\psi_i^{\mathrm{(ex)}} \in \mathcal{H}_i^{\mathrm{(ex)}} = \mathbb{R}^{d_i^{\mathrm{(ex)}}}
\label{eq:psi_ex}
\end{equation}
with dimension $d_i^{\mathrm{(ex)}}=2$ for all sites, i.e., only the ground and the lowest excited state of each site  $i$ are considered. 
Similarly, using the second quantization introduced in Eq.~\eqref{eq:H_ph2} of Sec.~\ref{sec:model}, a single site vibrational state vector is given by
\begin{equation}
\psi_i^{\mathrm{(ph)}} \in \mathcal{H}_i^{\mathrm{(ph)}} = \mathbb{R}^{d_i^{\mathrm{(ph)}}}
\label{eq:psi_ph}
\end{equation}
where in practice the dimension $d_i^{\mathrm{(ph)}}$ has to be determined by truncation to a suitable value, see our remarks in Sec.~\ref{sec:results_gen}. 
Hence, a coupled electronic-vibrational state vector of a single site will be given by a vector in the product space
\begin{equation}
\psi_i \in \mathcal{H}_i := \mathcal{H}_i^{\mathrm{(ex)}} \otimes \mathcal{H}_i^{\mathrm{(ph)}} \cong \mathbb{R}^{d_i}
\label{eq:psi_coup}
\end{equation}
with dimension $d_i:=d_i^{\mathrm{(ex)}}d_i^{\mathrm{(ph)}}$
Finally, a quantum state of the complete chain (or ring) comprising $N$ sites can be expressed as a tensor, i.e., a multidimensional array
\begin{equation}
\Psi  \in 
\mathcal{H} := \mathcal{H}_1 \otimes \mathcal{H}_2 \otimes \cdots \otimes \mathcal{H}_N \cong 
\mathbb{R}^{D}
\label{eq:psi_total}
\end{equation}
with a total dimension of $D:=\prod_{i=1}^N d_i$ where $N$ is called the order of the tensor. 
In order to obtain quantum states which are stationary solutions of the Hamiltonian \eqref{eq:H_total}, we need to solve the corresponding time-independent Schr\"{o}dinger equation (TISE), $H \Psi =E \Psi$, where $H$ is the tensorized version of \eqref{eq:H_total}, i.e., the numerical discretization of \eqref{eq:H_total} reshaped to a tensor in $\mathbb{R}^{(d_1 \times d_1) \times \dots \times (d_N \times d_N)}$. 
As a consequence, we only consider tensors with real-valued entries since $H$ is a symmetric operator which has an entirely real spectrum.
Here and in what follows, we use a simple matrix-like notation for the multiplication of tensors. For example, $H \Psi$ defines a tensor in $\mathbb{R}^{d_1 \times \dots \times d_N}$ with entries $(H \Psi)_{x_1, \dots, x_N} = \sum_{y_1 = 1}^{d_1} \dots \sum_{y_N= 1}^{d_N} H_{x_1, y_1, \dots, x_N, y_N} \Psi_{y_1, \dots, y_N}$.

In general, the solution of such eigenvalue problems in high-dimensional tensor spaces is extremely challenging or even impossible.
That is, the storage consumption of (non-sparse) tensors of the form \eqref{eq:H_total} or \eqref{eq:psi_total} grows exponentially with the order. 
Due to this curse of dimensionality, storing the considered tensors in full format may be infeasible for sufficiently large $N$. 
Therefore, we require special tensor representations and properly designed numerical methods, which we will introduce in the following sections.

\subsection{Tensor trains}\label{sec: tensor trains}

\begin{figure}[htbp]
\centering
\begin{subfigure}[b]{0.19\textwidth}
\centering
\begin{tikzpicture}
\draw[black] (0,0) -- node [label={[shift={(0,0.2)}]$d_1$}] {} ++ (0,0.7) ;
\draw[black] (0,0) -- node [label={[shift={(0.6,-0.2)}]$d_2$}] {} ++ (0.666,0.216) ;
\draw[black] (0,0) -- node [label={[shift={(0.4,-0.9)}]$d_3$}] {} ++ (0.411,-0.566) ;
\draw[black] (0,0) -- node [label={[shift={(-0.4,-0.9)}]$d_4$}] {} ++ (-0.411,-0.566) ;
\draw[black] (0,0) -- node [label={[shift={(-0.6,-0.2)}]$d_5$}] {} ++ (-0.666,0.216) ;
\node[draw,shape=circle,fill=Gray, scale=0.65] at (0,0){};
\end{tikzpicture}
\caption{}
\end{subfigure}
\hfill
\begin{subfigure}[b]{0.39\textwidth}
\centering
\begin{tikzpicture}
\draw[black] (0,0) -- node [label={[shift={(0,-0.15)}]$r_1$}] {} ++ (1,0) ;
\draw[black] (1,0) -- node [label={[shift={(0,-0.15)}]$r_2$}] {} ++ (1,0) ;
\draw[black] (2,0) -- node [label={[shift={(0,-0.15)}]$r_3$}] {} ++ (1,0) ;
\draw[black] (3,0) -- node [label={[shift={(0,-0.15)}]$r_4$}] {} ++ (1,0) ;
\draw[black] (0,0) -- node [label={[shift={(0,-1.1)}]$d_1$}] {} ++ (0,-0.7) ;
\draw[black] (1,0) -- node [label={[shift={(0,-1.1)}]$d_2$}] {} ++ (0,-0.7) ;
\draw[black] (2,0) -- node [label={[shift={(0,-1.1)}]$d_3$}] {} ++ (0,-0.7) ;
\draw[black] (3,0) -- node [label={[shift={(0,-1.1)}]$d_4$}] {} ++ (0,-0.7) ;
\draw[black] (4,0) -- node [label={[shift={(0,-1.1)}]$d_5$}] {} ++ (0,-0.7) ;
\node[draw,shape=circle,fill=Gray, scale=0.65] at (0,0){};
\node[draw,shape=circle,fill=Gray, scale=0.65] at (1,0){};
\node[draw,shape=circle,fill=Gray, scale=0.65] at (2,0){};
\node[draw,shape=circle,fill=Gray, scale=0.65] at (3,0){};
\node[draw,shape=circle,fill=Gray, scale=0.65] at (4,0){};
\end{tikzpicture}
\caption{}
\end{subfigure}
\hfill
\begin{subfigure}[b]{0.39\textwidth}
\centering
\begin{tikzpicture}
\draw[black] (0,0) -- node [label={[shift={(0,-0.15)}]$r_1$}] {} ++ (1,0) ;
\draw[black] (1,0) -- node [label={[shift={(0,-0.15)}]$r_2$}] {} ++ (1,0) ;
\draw[black] (2,0) -- node [label={[shift={(0,-0.15)}]$r_3$}] {} ++ (1,0) ;
\draw[black] (3,0) -- node [label={[shift={(0,-0.15)}]$r_4$}] {} ++ (1,0) ;
\draw[black] (0,0) -- node [label={[shift={(0,-1.1)}]$d_1$}] {} ++ (0,-0.7) ;
\draw[black] (1,0) -- node [label={[shift={(0,-1.1)}]$d_2$}] {} ++ (0,-0.7) ;
\draw[black] (2,0) -- node [label={[shift={(0,-1.1)}]$d_3$}] {} ++ (0,-0.7) ;
\draw[black] (3,0) -- node [label={[shift={(0,-1.1)}]$d_4$}] {} ++ (0,-0.7) ;
\draw[black] (4,0) -- node [label={[shift={(0,-1.1)}]$d_5$}] {} ++ (0,-0.7) ;
\draw[black] (0,0) -- node [label={[shift={(0,0.2)}]$d_1$}] {} ++ (0,0.7) ;
\draw[black] (1,0) -- node [label={[shift={(0,0.2)}]$d_2$}] {} ++ (0,0.7) ;
\draw[black] (2,0) -- node [label={[shift={(0,0.2)}]$d_3$}] {} ++ (0,0.7) ;
\draw[black] (3,0) -- node [label={[shift={(0,0.2)}]$d_4$}] {} ++ (0,0.7) ;
\draw[black] (4,0) -- node [label={[shift={(0,0.2)}]$d_5$}] {} ++ (0,0.7) ;
\node[draw,shape=circle,fill=Gray, scale=0.65] at (0,0){};
\node[draw,shape=circle,fill=Gray, scale=0.65] at (1,0){};
\node[draw,shape=circle,fill=Gray, scale=0.65] at (2,0){};
\node[draw,shape=circle,fill=Gray, scale=0.65] at (3,0){};
\node[draw,shape=circle,fill=Gray, scale=0.65] at (4,0){};
\end{tikzpicture}
\caption{}
\end{subfigure}
\caption{Graphical representation of tensors and tensor trains: Tensors are depicted as circles with different arms indicating the set of free indices. (a) Tensor of order $5$ in full format, e.g., a quantum state $\Psi$. (b) Tensor of order 5 in TT format, the first and the last core are matrices, the other cores are tensors of order 3. (c) Linear operator in TT format, the first and the last core are tensors of order 3, the other cores are tensors of order 4. Here we only consider Hamiltonians in $\mathbb{R}^{(d_1 \times d_1) \times (d_2 \times d_2) \times \dots \times (d_N \times d_N)}$, but complex operators are also possible.}
\label{fig: tensor trains}
\end{figure}
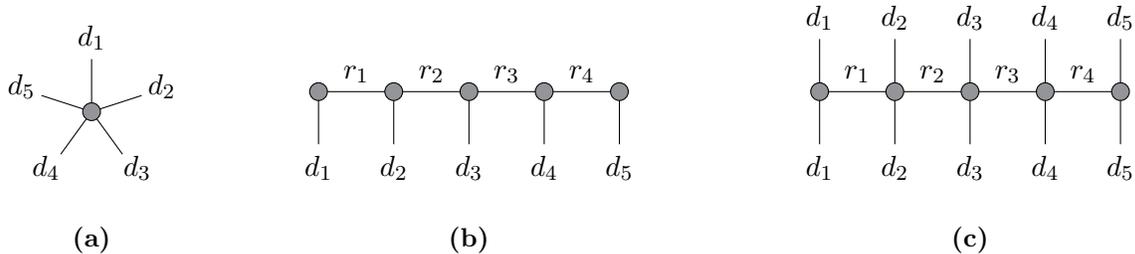

In terms of storage consumption and computational robustness, the \emph{tensor-train format} (TT format)~\cite{Oseledets2009a, Oseledets2009b} is a promising candidate for representing high-dimensional tensors as contractions of multiple low-dimensional tensors. 
The idea is to decompose a tensor of order $N$ -- in particular, a quantum state $\Psi$ -- into $N$ component tensors which are coupled in a chain, see Fig.~\ref{fig: tensor trains} and Def.~\ref{def: TT}.

\begin{definition}\label{def: TT}
 A tensor $\Psi \in \mathbb{R}^{d_1 \times d_2 \times \dots \times d_N}$ is said to be in the TT format if 
 \begin{equation}
  \Psi = \sum_{k_0=1}^{r_0} \cdots \sum_{k_N=1}^{r_N} \Psi^{(1)}_{k_0, :, k_1} \otimes \cdots \otimes \Psi^{(N)}_{k_{N-1}, :, k_N}.
  \label{eq:TT_def}
 \end{equation}
 The tensors $\Psi^{(i)} \in \mathbb{R}^{r_{i-1} \times d_i \times r_i}$ of order $3$ are called TT cores and the numbers $r_i$ are called TT ranks. It holds that $r_0=r_N = 1$ and $r_i \geq 1$ for $i = 1, \dots, N-1$.
 Any element of $\Psi$ can then be written as
 \begin{equation*}
  \Psi_{i_1, \dots , i_N} = \Psi^{(1)}_{:, i_1, :} \cdot \ldots \cdot \Psi^{(N)}_{:, i_N, :}.
 \end{equation*}
\end{definition}

In general, any tensor can be expressed in the TT format if we do not restrict the TT ranks, e.g., by using higher-order singular value decompositions~(HOSVD)~\cite{Oseledets2011}, cf.~Appendix~\ref{app: orthonormalization}. 
However, the TT ranks not only have a strong influence on the capability of representing a given tensor as a tensor train but also determine its storage consumption.
The storage consumption of a tensor $\Psi \in \mathbb{R}^{d_1 \times d_2 \times \dots \times d_N}$ in TT format can be estimated as $\mathcal{O}(\sum_{i=1}^N d_i \cdot r^2)$, where $r$ is the maximum of all TT ranks of $\Psi$.
Thus, we are particularly interested in low-rank TT decompositions because we are only able to mitigate the curse of dimensionality as long as the problem at hand allows for an acceptable accuracy of the approximate solution when we restrict ourselves to TT cores with ranks of manageable size.
 
Besides the graphical representation of tensor trains, we will use another notation at some points in this work in order to derive compact representations of tensor-train
operators. 
In this core notation, cf.~\cite{Kazeev2013, Gelss2017}, a TT core is represented as a two-dimensional array containing vectors or matrices as elements, respectively. 
For instance, the cores $\Psi^{(i)} \in \mathbb{R}^{r_{i-1} \times d_i \times r_i}$ of a given tensor $\Psi$ are written as
\begin{equation}
 \llbracket \Psi^{(i)} \rrbracket = 
 \left\llbracket
 \begin{matrix}
  \Psi^{(i)}_{1,:,1} & \cdots & \Psi^{(i)}_{1,:,r_i} \\
  \vdots                     & \ddots & \vdots                       \\
  \Psi^{(i)}_{r_{i-1},:,1} & \cdots & \Psi^{(i)}_{r_{i-1},:,r_i}
 \end{matrix}
 \right\rrbracket.
\end{equation}

\begin{example}
 Consider the tensor $\Psi \in \mathbb{R}^{2 \times 2 \times 2}$ defined by 
 \begin{equation*}
  \Psi_{:,:,1} = \begin{pmatrix} 0 & 2 \\ 1 & 3\end{pmatrix} \quad \text{and} \quad \Psi_{:,:,2} = \begin{pmatrix} 4 & 6 \\ 5& 7\end{pmatrix}.
 \end{equation*}
 In terms of tensor products, i.e., in canonical format, $\Psi$ can be written as
 \begin{equation*}
 \begin{split}
  \Psi &= \begin{pmatrix} 0 & 2 \\ 1 & 3\end{pmatrix} \otimes \begin{pmatrix}1 \\ 0 \end{pmatrix} + \begin{pmatrix} 4 & 6 \\ 5 & 7\end{pmatrix} \otimes \begin{pmatrix}0 \\ 1 \end{pmatrix} = \begin{pmatrix} 0 & 2 \\ 1 & 3\end{pmatrix} \otimes \begin{pmatrix}1 \\ 1 \end{pmatrix} + \begin{pmatrix} 4 & 4 \\ 4 & 4\end{pmatrix} \otimes \begin{pmatrix}0 \\ 1 \end{pmatrix} \\
  &= \begin{pmatrix} 0 \\ 1\end{pmatrix} \otimes \begin{pmatrix}1 \\ 1 \end{pmatrix} \otimes \begin{pmatrix}1 \\ 1 \end{pmatrix} + \begin{pmatrix} 1 \\ 1\end{pmatrix} \otimes \begin{pmatrix}0 \\ 2 \end{pmatrix} \otimes \begin{pmatrix}1 \\ 1 \end{pmatrix} + \begin{pmatrix} 1 \\ 1\end{pmatrix} \otimes \begin{pmatrix}1 \\ 1 \end{pmatrix} \otimes \begin{pmatrix}0 \\ 4 \end{pmatrix}.
 \end{split}
 \end{equation*}
 Using the core notation of the TT format, $\Psi$ can be expressed as 
 \begin{equation*}
  \Psi = \left\llbracket \begin{matrix} \begin{pmatrix} 0 \\ 1\end{pmatrix} & \begin{pmatrix} 1 \\ 1\end{pmatrix}
 \end{matrix}
 \right\rrbracket \otimes \left\llbracket \begin{matrix} \begin{pmatrix} 1 \\ 1\end{pmatrix} & \begin{pmatrix} 0 \\ 0\end{pmatrix} \\[0.5cm] \begin{pmatrix} 0 \\ 2\end{pmatrix} & \begin{pmatrix} 1 \\ 1\end{pmatrix}
 \end{matrix}
 \right\rrbracket \otimes \left\llbracket \begin{matrix} \begin{pmatrix} 1 \\ 1\end{pmatrix} \\[0.5cm] \begin{pmatrix} 0 \\ 4\end{pmatrix}
 \end{matrix}
 \right\rrbracket.
 \end{equation*}
 Note that the canonical rank of $\Psi$ is three while the TT rank is two.
\end{example}

In this work, we are not interested in decomposing a tensor into the TT format but rather want to approximate quantum states directly as tensor trains. 
The methods described in the following sections will use an orthonormalization procedure similar to HOSVD either to give the TT approximation a certain structure or to reduce the TT ranks, see Appendix~\ref{app: orthonormalization}.

\subsection{SLIM decomposition}\label{sec: SLIM}

In the quantum-mechanical Hamiltonians for coupled excitons and phonons introduced in Sec.~\ref{sec:model}, all terms are either acting locally on single sites or are of NN type coupling nearest neighbor sites. 
As shown in \cite{Gelss2017}, the structure of such a system corresponds to the topology of the TT format using a so-called \emph{SLIM decomposition}. 
That is, the \emph{canonical representation} of the Hamiltonian tensor $H \in \mathbb{R}^{(d_1 \times d_1) \times \dots \times (d_N \times d_N)}$ would only consist of elementary tensors, where at most two (adjacent) components are unequal to the identity matrix: 
\begin{equation}
    \begin{split}
        H &= S_{1} \otimes I_2 \otimes \dots \otimes I_N \quad + \quad \dots \quad + \quad I_1 \otimes \dots \otimes I_{N-1} \otimes S_{N} \\
        & \quad + \quad \sum_{\lambda=1}^{\xi_1}  L_{1,\lambda} \otimes M_{2,\lambda} \otimes I_3 \otimes \dots \otimes I_N \quad + \quad \dots \\
				& \quad + \quad \sum_{\lambda=1}^{\xi_{N-1}}  I_1 \otimes \dots \otimes I_{N-2} \otimes L_{N-1,\lambda} \otimes M_{N,\lambda}\\
        & \quad + \quad \sum_{\lambda=1}^{\xi_N} M_{1,\lambda} \otimes I_2 \otimes \dots \otimes I_{N-1} \otimes L_{N,\lambda}, 
    \end{split}
		\label{eq:SLIM_1}
\end{equation}
where all components $S_{i}$, $L_{i,\lambda}$, and $M_{i,\lambda}$ as well as the identities $I_i$ are matrices in $\mathbb{R}^{d_i \times d_i}$. 
The origin of the abbreviation SLIM is explained by the quantities above.
Note that the last line of Eq.~(\ref{eq:SLIM_1}) is only to comply with periodic boundary conditions of cyclic systems and can be omitted otherwise. 
The single site contributions to the total Hamiltonian, $S_i$, are derived from Eqs.~(\ref{eq:H_ex},\ref{eq:H_ph2},\ref{eq:H_couple_chi},\ref{eq:H_couple_rho}) yielding
\begin{equation}
	S_i = \alpha_i b_i^\dagger b_i 
	+ \tilde{\nu}_i \left(c_i^\dagger c_i + \frac{1}{2}\right)
	+ \left( \tilde{\chi}_i - \tilde{\rho}_i \right) 
	b_i^\dagger b_i \otimes \left( c_i^\dagger+c_i \right)
\label{eq:SLIM_S}
\end{equation}
The two-site contributions to the total Hamiltonian, see Eqs.~(\ref{eq:H_ex},\ref{eq:H_ph2},\ref{eq:H_couple_rho},\ref{eq:H_couple_sig},\ref{eq:H_couple_tau}), are expressible as sums of $\xi_i=9$ products of operators $L_{i,\lambda}$ acting on site $i$ and operators $M_{i+1, \lambda}$ acting on neighboring site $i+1$:
\begin{equation}
\begin{tabular}{lcl}
 $L_{i,1} = \beta_i b_i^\dagger$, & \hspace{3cm} & $M_{i+1,1} = b_{i+1}$, \\
 $L_{i,2} = \beta_i b_i$, &  & $M_{i+1,2} = b_{i+1}^\dagger$,\\
 $L_{i,3} = - \tilde{\omega}_i\left( c_i^\dagger+c_i \right)$, & & $M_{i+1,3} = c_{i+1}^\dagger+c_{i+1} $, \\
 $L_{i,4} = \left( \tilde{\rho}_i + \tilde{\sigma}_i \right) b_i^\dagger b_i$, & & $M_{i+1,4} = c_{i+1}^\dagger+c_{i+1} $, \\
 $L_{i,5} = - \left( c_i^\dagger+c_i \right)$, & & $M_{i+1,5} = \tilde{\sigma}_{i+1} b_{i+1}^\dagger b_{i+1}$, \\
 $L_{i,6} = +\tilde{\tau}_i b_i^\dagger$, & & $M_{i+1,6} = b_{i+1}\otimes\left( c_{i+1}^\dagger+c_{i+1} \right)$, \\
 $L_{i,7} = -\tilde{\tau}_i b_i^\dagger\otimes\left( c_i^\dagger+c_i \right)$, & & $M_{i+1,7} = b_{i+1}$, \\
 $L_{i,8} = +\tilde{\tau}_i b_i$, & & $M_{i+1,8} = b_{i+1}^\dagger\otimes\left( c_{i+1}^\dagger+c_{i+1} \right)$, \\
 $L_{i,9} = -\tilde{\tau}_i b_i\otimes\left( c_i^\dagger+c_i \right)$, & & $M_{i+1,9} = b_{i+1}^\dagger$.
\end{tabular}
\label{eq:SLIM_LM}
\end{equation}
Obviously, certain simplifications arise if some of the coefficients $\chi_i$, $\rho_i$, $\sigma_i$, or $\tau_i$ vanish, typically leading to a reduced number $\xi_i$. 
Gathering all components in corresponding core elements $S_i$, $L_i$, $I_i, M_i$, and $J_i$, see Appendix~\ref{app: SLIM supercores}, allows to express the linear operator $H$ as a SLIM decomposition given by
\begin{equation}
\begin{split}
    H & =  
    \left\llbracket\begin{matrix}
        S_1 & L_1 & I_1 & M_1 
    \end{matrix}\right\rrbracket
    \otimes 
    \left\llbracket\begin{matrix}
        I_2 & 0            & 0            & 0            \\
        M_2 & 0            & 0            & 0            \\       
        S_2 & L_2 & I_2 & 0            \\
        0            & 0            & 0            & J_2
    \end{matrix}\right\rrbracket
    \otimes \dots \\
		& \qquad \dots  \otimes
    \left\llbracket\begin{matrix}
        I_{N-1} & 0                & 0                & 0                \\
        M_{N-1} & 0                & 0                & 0                \\       
        S_{N-1} & L_{N-1} & I_{N-1} & 0                \\
        0                & 0                & 0                & J_{N-1}
    \end{matrix}\right\rrbracket
    \otimes
    \left\llbracket\begin{matrix}
        I_{N} \\
        M_{N} \\       
        S_{N} \\
        L_{N}
    \end{matrix}\right\rrbracket.
    \end{split}
		\label{eq:SLIM_3}
\end{equation}
The above equation holds for all heterogeneous, cyclic systems, a proof of which can be found in the Appendix of Ref.~\cite{Gelss2017}. 
For homogeneous systems where the interactions do not depend on the site indices, the core elements $S_i$, $L_i$, $I_i$, $M_i$, and $J_i$ do not change for different $i$. 
Thus, the TT ranks of the supercores are fixed for an increasing number of sites resulting in a linear growth of the storage consumption. 
For heterogeneous systems the same argumentation holds if we assume the TT ranks to be bounded, i.e., the number of two-site contributions is bounded. 
Due to the repeating pattern of the SLIM decomposition, we only have to insert or remove certain TT cores, respectively, if we want to increase or decrease the number of sites. 

The efficiency of the tensor-based algorithms described in the following sections depends strongly on the TT ranks of the operator $H$. 
Thus, the aim is to mitigate the curse of dimensionality by using low-rank
TT representations of the Hamiltonian tensor as well as of the quantum states.

\subsection{Alternating linear scheme}
\label{sec: ALS}

Eigenvalue problems in TT format can be approached by different methods. 
One choice are iterative algorithms based on the multiplication/contraction and truncation, cf.~Appendix~\ref{app: orthonormalization}, of tensor trains. 
For instance, \emph{power iteration}~\cite{Klus2016} and \emph{Lanczos methods}~\cite{Nip2013} can be used to approximate a set of dominant eigenvalues and corresponding eigentensors.
However, the performance of both approaches may suffer from the numerical complexity of repeated tensor truncations, particularly the latter one involves the computation of an uncertain number of Lanczos vectors (represented in TT format) needed for the approximation of the eigentensors.

In this work, we will rely on an established algorithm for solving optimization problems such as systems of linear equations and eigenvalue problems in the TT format, the so-called \emph{alternating linear scheme} (ALS), see \cite{Holtz2012}, which is strongly related to the density matrix renormalization group algorithm~\cite{White1992}.
ALS is based on the alternating optimization of the TT cores of a given initial guess and has been successfully applied to various high-dimensional eigenvalue problems, see, e.g., Refs.~\cite{Dolgov2014, Gelss2016, Benner2017}. 
This method does not involve the truncation of TT decompositions since, by design, ALS operates only on a manifold of tensor trains with fixed ranks. 
Nevertheless, we will use the decomposition technique described in Appendix~\ref{app: orthonormalization} to ensure the stability of the scheme.
 
Consider an eigenvalue problem of the form
\begin{equation}\label{eq: EVP}
 H \Psi = E \Psi, \quad \Psi \neq 0,
\end{equation}
which is given in TT format with tensor trains $H \in \mathbb{R}^{(d_1 \times d_1) \times \dots \times (d_N \times d_N)}$ and $\Psi \in \mathbb{R}^{d_1 \times \dots \times d_N}$. 
The solutions of Eq.~\eqref{eq: EVP} are given by the stationary points of the \emph{Rayleigh quotient}
\begin{equation}\label{eq: Rayleigh quotient}
 \mathcal{R}(H, \Psi) = \frac{\Psi^\top H \Psi}{\Psi^\top \Psi}.
\end{equation}
In order to compute, e.g., the smallest eigenvalue of $H$, we want to find the minimizer of $\mathcal{R}$. But instead of computing this minimizer in a single step, which may be infeasible for high-dimensional tensors, the idea of ALS is to optimize the TT cores successively by restricting the Rayleigh quotient to single dimensions. 
That is, given an initial guess of the solution of Eq.~\eqref{eq: EVP}, the ALS algorithm updates the TT cores iteratively by constructing and solving low-dimensional eigenvalue problems during repeated bidirectional half sweeps.

For the construction of the subsystems for each iteration step, all cores of the solution which are not optimized at this point are contracted with the TT cores of the operator $H$, see Fig.~\ref{fig:ALS}.
The (reshaped) solution is then either left-orthonormalized (first half sweep) or right-orthonormalized (second half sweep), see Appendix~\ref{app: orthonormalization}.
An important requirement of ALS is that the considered TT operator must be symmetric. 
If it is additionally positive definite, we know that the condition number of the subsystems is bounded by the condition number of the operator.

\begin{figure}[htbp]
\centering
\begin{tikzpicture}

\draw[draw=Gray, dashed] (-2,0.5) --++ (8.5,0);
\draw[draw=Gray, dashed] (-2,-0.5) --++ (8.5,0);
\node[anchor=west] at (-2,1) {\textcolor{Gray}{$Q_i$}};
\node[anchor=west] at (-2,0) {\textcolor{Gray}{$H$}};
\node[anchor=west] at (-2,-1) {\textcolor{Gray}{$Q_i^\top$}};
\draw[] (0,1) --++ (0.66,0);
\draw[dotted] (0.66,1) --++ (0.66,0);
\draw[] (1.33,1) --++ (1.33,0);
\draw[] (3.33,1) --++ (1.33,0);
\draw[dotted] (4.66,1) --++ (0.66,0);
\draw[] (5.33,1) --++ (0.66,0);
\draw[] (0,0) --++ (0.66,0);
\draw[dotted] (0.66,0) --++ (0.66,0);
\draw[] (1.33,0) --++ (3.33,0);
\draw[dotted] (4.66,0) --++ (0.66,0);
\draw[] (5.33,0) --++ (0.66,0);
\draw[] (0,-1) --++ (0.66,0);
\draw[dotted] (0.66,-1) --++ (0.66,0);
\draw[] (1.33,-1) --++ (1.33,0);
\draw[] (3.33,-1) --++ (1.33,0);
\draw[dotted] (4.66,-1) --++ (0.66,0);
\draw[] (5.33,-1) --++ (0.66,0);
\draw[] (0,1) --++ (0,-2);
\draw[] (2,1) --++ (0,-2);
\draw[] (3,0.66) --++ (0,-1.33);
\draw[] (4,1) --++ (0,-2);
\draw[] (6,1) --++ (0,-2);

\draw[] (3,2) --++ (0,-0.33);
\draw[] (2.66,2) --++ (0.66,0);
\node[draw,shape=circle,fill=Orange, scale=0.65] at (3,2){};
\draw[Gray, ->, >=latex] (3,1.46) --++ (0,-0.6);
\draw[Gray, ->, >=latex] (2.66,1.8) --++ (0,-0.6);
\draw[Gray, ->, >=latex] (3.33,1.8) --++ (0,-0.6);

\node[draw,shape=semicircle,rotate=135,fill=white, anchor=south,inner sep=2pt, outer sep=0pt, scale=0.75] at (0,1){}; 
\node[draw,shape=semicircle,rotate=315,fill=Blue, anchor=south,inner sep=2pt, outer sep=0pt, scale=0.75] at (0,1){};
\node[draw,shape=semicircle,rotate=135,fill=white, anchor=south,inner sep=2pt, outer sep=0pt, scale=0.75] at (2,1){}; 
\node[draw,shape=semicircle,rotate=315,fill=Blue, anchor=south,inner sep=2pt, outer sep=0pt, scale=0.75] at (2,1){};
\node[draw,shape=semicircle,rotate=225,fill=white, anchor=south,inner sep=2pt, outer sep=0pt, scale=0.75] at (4,1){}; 
\node[draw,shape=semicircle,rotate=45,fill=Blue, anchor=south,inner sep=2pt, outer sep=0pt, scale=0.75] at (4,1){};
\node[draw,shape=semicircle,rotate=225,fill=white, anchor=south,inner sep=2pt, outer sep=0pt, scale=0.75] at (6,1){}; 
\node[draw,shape=semicircle,rotate=45,fill=Blue, anchor=south,inner sep=2pt, outer sep=0pt, scale=0.75] at (6,1){};
\node[draw,shape=circle,fill=Green, scale=0.65] at (0,0){};
\node[draw,shape=circle,fill=Green, scale=0.65] at (2,0){};
\node[draw,shape=circle,fill=Green, scale=0.65] at (3,0){};
\node[draw,shape=circle,fill=Green, scale=0.65] at (4,0){};
\node[draw,shape=circle,fill=Green, scale=0.65] at (6,0){};
\node[draw,shape=semicircle,rotate=45,fill=white, anchor=south,inner sep=2pt, outer sep=0pt, scale=0.75] at (0,-1){}; 
\node[draw,shape=semicircle,rotate=225,fill=Blue, anchor=south,inner sep=2pt, outer sep=0pt, scale=0.75] at (0,-1){};
\node[draw,shape=semicircle,rotate=45,fill=white, anchor=south,inner sep=2pt, outer sep=0pt, scale=0.75] at (2,-1){}; 
\node[draw,shape=semicircle,rotate=225,fill=Blue, anchor=south,inner sep=2pt, outer sep=0pt, scale=0.75] at (2,-1){};
\node[draw,shape=semicircle,rotate=315,fill=white, anchor=south,inner sep=2pt, outer sep=0pt, scale=0.75] at (4,-1){}; 
\node[draw,shape=semicircle,rotate=135,fill=Blue, anchor=south,inner sep=2pt, outer sep=0pt, scale=0.75] at (4,-1){};
\node[draw,shape=semicircle,rotate=315,fill=white, anchor=south,inner sep=2pt, outer sep=0pt, scale=0.75] at (6,-1){}; 
\node[draw,shape=semicircle,rotate=135,fill=Blue, anchor=south,inner sep=2pt, outer sep=0pt, scale=0.75] at (6,-1){};
\end{tikzpicture}
\caption{Construction of the ALS subsystems: The contraction of the operator $H$ (green circles) and the fixed cores of the current solution $\Psi$ (blue circles, also called retraction operator $Q_i$) is represented by joining corresponding arms. The orthonormality of the TT cores of of $\Psi$, which is required for the stability of ALS, is depicted by half-filled circles, cf.~Fig.~\ref{fig: ortho}. The result is an order-$6$ tensor that, reshaped into a matrix, gives the operator of the low-dimensional subsystem. Depending on the problem, the updated TT core $\Psi^{(i)}$ (orange circle) can be obtained by either solving a system of linear equations or -- as in our case -- an eigenvalue problem, respectively.}
\label{fig:ALS}
\end{figure}

ALS uses an efficient stacking method for the contractions of the fixed TT cores of $\Psi$ and the operator $H$ while iteratively updating the solution.
Applying the retraction operator $Q_i$, see Fig.~\ref{fig:ALS}, leads to subproblems of the form (in simplified notation)
\begin{equation}\label{eq:subsystems}
( Q_i^\top H Q_i) \Psi^{(i)} = E Q_i^\top \Psi  = E \Psi^{(i)}
\end{equation}
For a detailed explanation on how to construct the operator as well as the right-hand sides for optimization problems in TT format, we refer to~\cite{Holtz2012}.

Suppose the TT operator (\ref{eq:SLIM_3}) with order $N$ (number of sites) and maximum mode size $d$ (dimension of the Hilbert space from Eq.~(\ref{eq:psi_coup})) has maximum rank $R$. 
The computational complexity of ALS can be divided into two parts, namely constructing the subsystems and solving the low-dimensional eigenvalue problems. 
Due to the involved tensor contractions needed to construct the matrix operators for all cores, the former one can be estimated as $\mathcal{O}(s N d^2 r^4 R^2)$, where $r$ is the maximum rank of the solution and $s$ is the number of ALS repeats. 
The numerical costs for solving the associated eigenvalue problems strongly depend on the used methods. 
In practice, the complexity of an eigenvalue decomposition of a $d r^2 \times d r^2$ matrix is estimated as $\mathcal{O}(d^3 r^6)$. 
However, since only extremal eigenvalues have to be considered during the computations, this estimate can be lowered -- in the best case to $\mathcal{O}(d^2 r^4)$. 
Thus, the overall computational complexity is substantially given by $\mathcal{O}(s N d^2 r^4 R^2)$. 

In the case that the TT ranks of the solution and the number of repeats $s$ do not increase with the order, the computational effort scales linearly with $N$ which is of key importance when treating large systems.
The ranks of the TT operator are naturally bounded due to the restriction to NN interactions only, which in turn has allowed us to formulate the Hamiltonian in TT format, see Eq.~\eqref{eq:SLIM_3}.
For instance, for homogeneous and periodic systems, $\xi_1 = \dots = \xi_N =: \xi$ and, therefore, $R = 2 + 2 \xi$, see Sec.~\ref{sec: SLIM}.

\begin{remark}
 The authors are aware that the computational costs of ALS can be reduced by exploiting iterative methods for solving eigenvalue problems directly on the tensor network (without constructing the matrices for the subsystems explicitly). 
 As described in, e.g., \cite{Holtz2012}, the complexity depends only cubically on the rank $r$ if we compute the contraction shown in Fig.~\ref{fig:ALS} in an optimal way. However, the implementation even of simple iterative schemes compatible with the approach introduced in the following section would become highly complex.
 By constructing the subsystems explicitly in matrix form, we can rely on established software libraries including implementations of state-of-the-art eigenvalue algorithms.
\end{remark}

Note that ALS is designed to compute eigentensors corresponding to the smallest/largest magnitude eigenvalue~\cite{Holtz2012} of a given symmetric TT operator $H$. 
However, by determining the smallest eigenvalue of the shifted operator $H-\eta I$, $\eta \in \mathbb{R}$, we are able to compute eigentensors whose associated eigenvalues are close to $\eta$ since $(H - \eta I) \Psi = E_\text{shift} \Psi$ implies $H \Psi = (\eta + E_\text{shift}) \Psi$. From Eq.~\eqref{eq:subsystems} then follows that we can directly apply the eigenvalue shifting to the low-dimensional subproblems because orthonormality of $Q_i$ implies
\begin{equation}\label{eq: ev_shifting}
 ( Q_i^\top (H - \eta I) Q_i) \Psi^{(i)} = ( Q_i^\top H Q_i - \eta I) \Psi^{(i)} = E_\text{shift} \Psi^{(i)}.
\end{equation}
However, prior knowledge about the distribution of eigenvalues is needed for choosing approriate shifts $\eta$. 
Since we here are interested in the sequence of the $K$ smallest eigenvalues, an alternative method for the modification of the Hamiltonian is described in the following section.

\subsection{Integrated Wielandt deflation}
\label{subsec:IntegratedWielandt}

As described above, application of ALS to Eq.~\eqref{eq: EVP} allows us to directly obtain the smallest eigenvalue (energy) and the corresponding eigentensor (quantum state). 
As presented in \cite{Dolgov2014}, it is possible to approximate the $K$ smallest and largest eigenvalues, respectively, in one step by employing the so-called \emph{block TT format}. 
In general, this approach only works if the tensor trains representing the $K$ considered eigentensors can be expressed in a common block tensor train, i.e., all eigentensors can be written with the same TT cores except one. 
Our experiments showed that this is not the case for the eigenstates of the TISE. 
Thus, we propose another method for computing the leading eigentensors of an eigenvalue problem in TT format.
In what follows, we will focus on eigentensors corresponding to the smallest eigenvalues.
However, the approach can also be adapted to the computation of the largest eigenvalues and associated eigentensors.

To compute not only the first but rather the $K$ smallest eigenpairs, we modify the Hamiltonian after each computation, i.e., we want to displace the known eigenvalues, while keeping all other eigenvalues unchanged. 
For that purpose, we employ the \emph{Wielandt deflation} technique~\cite{Saad2011}.
Assuming we have calculated the first eigenpair $(E_0, \Psi_0)$ of $H_0 := H$, we define the first deflated Hamiltonian as
\begin{equation}
 H_1 = H_0 + \delta \Psi_0 \tilde{\Psi}_0^\top,
 \label{eq:deflated_H}
\end{equation}
where $\delta$ is a shift parameter such that $E_0$ is transformed into the eigenvalue $E_0 + \delta$.
In fact, the tensor $\tilde{\Psi}_0$ could be chosen arbitrarily as long as it satisfies $\tilde{\Psi}_0^\top \Psi_0 = 1$.
In standard linear algebra, two optimal choices for $\tilde{\Psi}_0$ in terms of the condition number of $H_{1}$ would be either a vector with all entries equal to one, rescaled to meet the normalization condition, or $\Psi_0$ itself. 
Since we want the deflated Hamiltonian to be again symmetric and positive definite, we choose $\tilde{\Psi}_0 = \Psi_0$.
Then, symmetry of $H_1$ implies that it has the same (mutually orthogonal) eigentensors as $H_0$ where only the eigenvalue corresponding to $\Psi_0$ is altered.
Note that $H_0$ can also be defined as a shifted operator of the form $H - \eta I$. 
In this case, we use the Wielandt deflation to approximate the $K$ eigenvalues closest to $\eta$, see Sec.~\ref{sec: ALS}. 

Repeating the above procedure, we have to solve the eigenvalue problems $H_k \Psi_k = E_k \Psi_k$, $k=0, \dots, K-1$ with TT operators of the form
\begin{equation}
H_{k} = H_{k-1} + \delta \Psi_{k-1} \Psi_{k-1}^\top = H_0 + \delta \sum_{j=0}^{k-1} \Psi_j \Psi_j^\top,
\label{eq:deflated_H_general}
\end{equation}
where $\Psi_j$ is the $j$th found eigentensor.
However, computing the outer product of $\Psi_j$ with itself may lead to a significant increase of the TT ranks of our new operator because the ranks of $ \Psi_j \Psi_j^\top $ are the squares of the ranks of $\Psi_j$. 
Therefore, we propose a method which exploits the deflation technique as given in Eq.~\eqref{eq:deflated_H_general} without explicitly constructing the TT operator $H_k$.

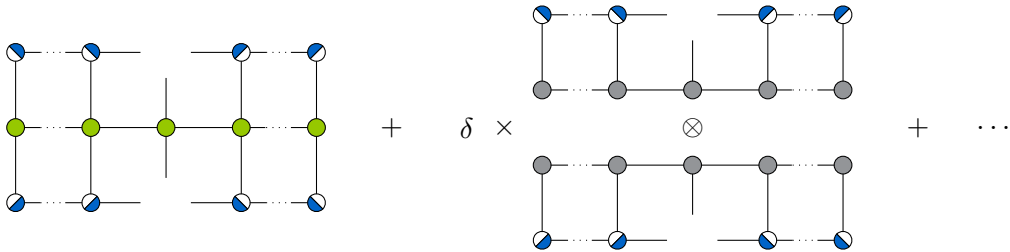
\begin{figure}[htbp]
	\centering
	\begin{tikzpicture}

	\draw[] (0,1) --++ (0.33,0);
	\draw[dotted] (0.33,1) --++ (0.33,0);
	\draw[] (0.66,1) --++ (1,0);
	\draw[] (2.33,1) --++ (1,0);
	\draw[dotted] (3.33,1) --++ (0.33,0);
	\draw[] (3.66,1) --++ (0.33,0);
	\draw[] (0,0) --++ (0.33,0);
	\draw[dotted] (0.33,0) --++ (0.33,0);
	\draw[] (0.66,0) --++ (2.66,0);
	\draw[dotted] (3.33,0) --++ (0.33,0);
	\draw[] (3.66,0) --++ (0.33,0);
	\draw[] (0,-1) --++ (0.33,0);
	\draw[dotted] (0.33,-1) --++ (0.33,0);
	\draw[] (0.66,-1) --++ (1,0);
	\draw[] (2.33,-1) --++ (1,0);
	\draw[dotted] (3.33,-1) --++ (0.33,0);
	\draw[] (3.66,-1) --++ (0.33,0);
	\draw[] (0,1) --++ (0,-2);
	\draw[] (1,1) --++ (0,-2);
	\draw[] (2,0.66) --++ (0,-1.33);
	\draw[] (3,1) --++ (0,-2);
	\draw[] (4,1) --++ (0,-2);

	\node[draw,shape=semicircle,rotate=135,fill=white, anchor=south,inner sep=2pt, outer sep=0pt, scale=0.75] at (0,1){}; 
	\node[draw,shape=semicircle,rotate=315,fill=Blue, anchor=south,inner sep=2pt, outer sep=0pt, scale=0.75] at (0,1){};
	\node[draw,shape=semicircle,rotate=135,fill=white, anchor=south,inner sep=2pt, outer sep=0pt, scale=0.75] at (1,1){}; 
	\node[draw,shape=semicircle,rotate=315,fill=Blue, anchor=south,inner sep=2pt, outer sep=0pt, scale=0.75] at (1,1){};
	\node[draw,shape=semicircle,rotate=225,fill=white, anchor=south,inner sep=2pt, outer sep=0pt, scale=0.75] at (3,1){}; 
	\node[draw,shape=semicircle,rotate=45,fill=Blue, anchor=south,inner sep=2pt, outer sep=0pt, scale=0.75] at (3,1){};
	\node[draw,shape=semicircle,rotate=225,fill=white, anchor=south,inner sep=2pt, outer sep=0pt, scale=0.75] at (4,1){}; 
	\node[draw,shape=semicircle,rotate=45,fill=Blue, anchor=south,inner sep=2pt, outer sep=0pt, scale=0.75] at (4,1){};
	\node[draw,shape=circle,fill=Green, scale=0.65] at (0,0){};
	\node[draw,shape=circle,fill=Green, scale=0.65] at (1,0){};
	\node[draw,shape=circle,fill=Green, scale=0.65] at (2,0){};
	\node[draw,shape=circle,fill=Green, scale=0.65] at (3,0){};
	\node[draw,shape=circle,fill=Green, scale=0.65] at (4,0){};
	\node[draw,shape=semicircle,rotate=45,fill=white, anchor=south,inner sep=2pt, outer sep=0pt, scale=0.75] at (0,-1){}; 
	\node[draw,shape=semicircle,rotate=225,fill=Blue, anchor=south,inner sep=2pt, outer sep=0pt, scale=0.75] at (0,-1){};
	\node[draw,shape=semicircle,rotate=45,fill=white, anchor=south,inner sep=2pt, outer sep=0pt, scale=0.75] at (1,-1){}; 
	\node[draw,shape=semicircle,rotate=225,fill=Blue, anchor=south,inner sep=2pt, outer sep=0pt, scale=0.75] at (1,-1){};
	\node[draw,shape=semicircle,rotate=315,fill=white, anchor=south,inner sep=2pt, outer sep=0pt, scale=0.75] at (3,-1){}; 
	\node[draw,shape=semicircle,rotate=135,fill=Blue, anchor=south,inner sep=2pt, outer sep=0pt, scale=0.75] at (3,-1){};
	\node[draw,shape=semicircle,rotate=315,fill=white, anchor=south,inner sep=2pt, outer sep=0pt, scale=0.75] at (4,-1){}; 
	\node[draw,shape=semicircle,rotate=135,fill=Blue, anchor=south,inner sep=2pt, outer sep=0pt, scale=0.75] at (4,-1){};
	
	\node[] at (5,0) {$+$};
	\node[] at (6,0) {$\delta$};
	\node[] at (6.5,0) {$\times$};
	
	\def\x{7}
	\def\y{0.5}
	\draw[] (0 + \x,1 + \y) --++ (0.33,0);
	\draw[dotted] (0.33 + \x,1 + \y) --++ (0.33,0);
	\draw[] (0.66 + \x,1 + \y) --++ (1,0);
	\draw[] (2.33 + \x,1 + \y) --++ (1,0);
	\draw[dotted] (3.33 + \x,1 + \y) --++ (0.33,0);
	\draw[] (3.66 + \x,1 + \y) --++ (0.33,0);
	\draw[] (0 + \x,0 + \y) --++ (0.33,0);
	\draw[dotted] (0.33 + \x,0 + \y) --++ (0.33,0);
	\draw[] (0.66 + \x,0 + \y) --++ (2.66,0);
	\draw[dotted] (3.33 + \x,0 + \y) --++ (0.33,0);
	\draw[] (3.66 + \x,0 + \y) --++ (0.33,0);
	\draw[] (0 + \x,1 + \y) --++ (0,-1);
	\draw[] (1 + \x,1 + \y) --++ (0,-1);
	\draw[] (2 + \x,0.66 + \y) --++ (0,-0.66);
	\draw[] (3 + \x,1 + \y) --++ (0,-1);
	\draw[] (4 + \x,1 + \y) --++ (0,-1);

	\node[draw,shape=semicircle,rotate=135,fill=white, anchor=south,inner sep=2pt, outer sep=0pt, scale=0.75] at (\x + 0,\y + 1){}; 
	\node[draw,shape=semicircle,rotate=315,fill=Blue, anchor=south,inner sep=2pt, outer sep=0pt, scale=0.75] at (\x + 0,\y + 1){};
	\node[draw,shape=semicircle,rotate=135,fill=white, anchor=south,inner sep=2pt, outer sep=0pt, scale=0.75] at (\x + 1,\y + 1){}; 
	\node[draw,shape=semicircle,rotate=315,fill=Blue, anchor=south,inner sep=2pt, outer sep=0pt, scale=0.75] at (\x + 1,\y + 1){};
	\node[draw,shape=semicircle,rotate=225,fill=white, anchor=south,inner sep=2pt, outer sep=0pt, scale=0.75] at (\x + 3,\y + 1){}; 
	\node[draw,shape=semicircle,rotate=45,fill=Blue, anchor=south,inner sep=2pt, outer sep=0pt, scale=0.75] at (\x + 3,\y + 1){};
	\node[draw,shape=semicircle,rotate=225,fill=white, anchor=south,inner sep=2pt, outer sep=0pt, scale=0.75] at (\x + 4,\y + 1){}; 
	\node[draw,shape=semicircle,rotate=45,fill=Blue, anchor=south,inner sep=2pt, outer sep=0pt, scale=0.75] at (\x + 4,\y + 1){};
	\node[draw,shape=circle,fill=Gray, scale=0.65] at (\x + 0,\y + 0){};
	\node[draw,shape=circle,fill=Gray, scale=0.65] at (\x + 1,\y + 0){};
	\node[draw,shape=circle,fill=Gray, scale=0.65] at (\x + 2,\y + 0){};
	\node[draw,shape=circle,fill=Gray, scale=0.65] at (\x + 3,\y + 0){};
	\node[draw,shape=circle,fill=Gray, scale=0.65] at (\x + 4,\y + 0){};
	
	\node[] at (\x+2,0) {$\otimes$};
	\def\x{7}
	\def\y{-1.5}
	\draw[] (0 + \x,\y) --++ (0.33,0);
	\draw[dotted] (0.33 + \x,\y) --++ (0.33,0);
	\draw[] (0.66 + \x,\y) --++ (1,0);
	\draw[] (2.33 + \x,\y) --++ (1,0);
	\draw[dotted] (3.33 + \x,\y) --++ (0.33,0);
	\draw[] (3.66 + \x,\y) --++ (0.33,0);
	\draw[] (0 + \x,1 + \y) --++ (0.33,0);
	\draw[dotted] (0.33 + \x,1 + \y) --++ (0.33,0);
	\draw[] (0.66 + \x,1 + \y) --++ (2.66,0);
	\draw[dotted] (3.33 + \x,1 + \y) --++ (0.33,0);
	\draw[] (3.66 + \x,1 + \y) --++ (0.33,0);
	\draw[] (0 + \x,1 + \y) --++ (0,-1);
	\draw[] (1 + \x,1 + \y) --++ (0,-1);
	\draw[] (2 + \x,1 + \y) --++ (0,-0.66);
	\draw[] (3 + \x,1 + \y) --++ (0,-1);
	\draw[] (4 + \x,1 + \y) --++ (0,-1);
	
	\node[draw,shape=circle,fill=Gray, scale=0.65] at (\x + 0,\y + 1){};
	\node[draw,shape=circle,fill=Gray, scale=0.65] at (\x + 1,\y + 1){};
	\node[draw,shape=circle,fill=Gray, scale=0.65] at (\x + 2,\y + 1){};
	\node[draw,shape=circle,fill=Gray, scale=0.65] at (\x + 3,\y + 1){};
	\node[draw,shape=circle,fill=Gray, scale=0.65] at (\x + 4,\y + 1){};
	
	\node[draw,shape=semicircle,rotate=45,fill=white, anchor=south,inner sep=2pt, outer sep=0pt, scale=0.75] at (\x + 0,\y){}; 
	\node[draw,shape=semicircle,rotate=225,fill=Blue, anchor=south,inner sep=2pt, outer sep=0pt, scale=0.75] at (\x + 0,\y){};
	\node[draw,shape=semicircle,rotate=45,fill=white, anchor=south,inner sep=2pt, outer sep=0pt, scale=0.75] at (\x + 1,\y){}; 
	\node[draw,shape=semicircle,rotate=225,fill=Blue, anchor=south,inner sep=2pt, outer sep=0pt, scale=0.75] at (\x + 1,\y){};
	\node[draw,shape=semicircle,rotate=315,fill=white, anchor=south,inner sep=2pt, outer sep=0pt, scale=0.75] at (\x + 3,\y){}; 
	\node[draw,shape=semicircle,rotate=135,fill=Blue, anchor=south,inner sep=2pt, outer sep=0pt, scale=0.75] at (\x + 3,\y){};
	\node[draw,shape=semicircle,rotate=315,fill=white, anchor=south,inner sep=2pt, outer sep=0pt, scale=0.75] at (\x + 4,\y){}; 
	\node[draw,shape=semicircle,rotate=135,fill=Blue, anchor=south,inner sep=2pt, outer sep=0pt, scale=0.75] at (\x + 4,\y){};
	
	\node[] at (\x+5,0) {$+$};
	\node[] at (\x+6,0) {$\ldots$};

	\end{tikzpicture}
	\caption{Construction of the deflated ALS subsystems: For each core optimization, the operator of the corresponding subsystem is the sum of the Hamiltonian (green circles) contracted with the retraction operator (blue circles) from both sides (cf.~Fig.~\ref{fig:ALS}) and a series of outer products of the form $\Theta_j \otimes \Theta_j$ scaled by the factor $\delta$, where $\Theta_j = Q_i^\top \Psi_j$ (reshaped as a vector, cores of $\Psi_j$ are depicted as gray circles) for $j=0, \dots, k-1$.}
	\label{fig: deflated_subproblem}
\end{figure}

In terms of the retraction operators $Q_i$, see Sec.~\ref{sec: ALS}, the subsystems which are solved during the ALS iterations are of the form
\begin{equation}\label{eq: deflated_subproblem_1}
  (Q_i^\top H_k Q_i) \Psi_k^{(i)} = E_k^{(i)} \Psi_k^{(i)},
\end{equation}
see Eq.~\eqref{eq:subsystems}. As for the global eigenvalue problem $H_k \Psi_k = E_k \Psi_k$, we are here also interested in finding the smallest eigenvalue $E_k^{(i)}$ -- or the eigenvalue closest to $\eta$, see Eq.~\eqref{eq: ev_shifting} -- and the corresponding eigenvector, represented by $\Psi^{(i)}_k$. 
As explained in the previous section, the retraction operator depends on the fixed cores of $\Psi_k$ in each iteration step of the ALS. 
If we now insert the deflated Hamiltonian tensor from Eq.~\eqref{eq:deflated_H_general} into Eq.~\eqref{eq: deflated_subproblem_1}, we obtain
\begin{equation}\label{eq: deflated_subproblem_2}
  (Q_i^\top H_0 Q_i + \delta \sum_{j=0}^{k-1} Q_i^\top \Psi_j \Psi_j^\top Q_i) \Psi^{(i)}_k = E^{(i)}_k \Psi^{(i)}_k.
\end{equation}
See Fig.~\ref{fig: deflated_subproblem} for a graphical representation of the operator on the left-hand side in Eq.~\eqref{eq: deflated_subproblem_2}.

Instead of constructing $H_k$ in TT format -- and thus facing a rapid increase of the TT ranks of the operators -- we directly manipulate the subsystems occuring in the ALS. 
That is, we only have to compute the tensors $\Theta_j = Q_i^\top \Psi_j$ for all known eigentensors $\Psi_j$ and add the matricizations of $\delta \Theta_j \otimes \Theta_j$ for $j = 0, \dots, k-1$ to the initial operator $Q_i^\top H_0 Q_i $. 
In comparison to the explicit construction of $H_k$, the proposed approach requires significantly less storage consumption and computational costs. The only dominant additional costs that arise (compared to standard ALS applied to $H_0$) result from the construction of the matricizations of $\delta \Theta_j \otimes \Theta_j$.
The computational complexity of the integrated deflation method can be estimated as $\mathcal{O} (s N (d^2 r^4 R^2 + k d^2 r^4)) = \mathcal{O} (s N d^2 r^4 R^2)$, cf.~Sec.~\ref{sec: ALS}.

Note that, in spite of the fixed TT ranks of the operator, the computational effort still increases for each computed eigenpair depending on the TT ranks of the previously computed eigentensors. 
However, the benefit in terms of performance is nevertheless remarkable compared to the application of ALS to the explicitly constructed $H_k$ given in Eq.~\eqref{eq:deflated_H_general}. Furthermore, we are able to handle the numerical costs for computing the eigenvalues of highest interest in practice if the quantum states can be approximated accurately enough by low-rank decompositions.
It turns out that, depending on the implementation, solving the low-dimensional eigenvalue problems for optimizing the TT cores of the solution must often be considered as the main bottleneck.
For growing $k$, the relative effect on the computational complexity of the Wielandt deflation often tends to decrease in practice.

\section{Results and Discussion}
\label{sec:results}
\subsection{Model parameters}
\label{sec:results_gen}

Here, we present and discuss typical results, aiming at assessing the performance and accuracy of the TT techniques to solve the TISE for excitons, phonons, and coupled systems.
Where applicable, we compare our numerical results with analytical solutions which are partly available for homogenous chains or rings, i.e., systems comprising $N$ identical sites.
Since we do not intend to model specific materials, a choice of parameters may suffice here which can be considered prototypical for excitonic energy transfer processes in organic semiconductors.
In atomic (Hartree) units, we choose a local excitation energy $\alpha=0.1 E_h$ ($\approx 2.7 $ eV) which is of the order of typical band gaps.
Next, we choose a NN coupling energy which is one order of magnitude smaller, $\beta=-0.01 E_h$ which can be regarded as typical for the situation in crystals of polyacene molecules.
Here the negative sign is chosen in analogy to a J--aggregate with head-to-tail alignment of the constituting molecules, which is typically found in these crystals~\cite{Hestand2017}.
For the description of the phonons we use dimensionless displacement coordinates with unit masses, $m=1$.
The harmonic frequencies are $\nu=10^{-3} E_h/\hbar$ ($\approx 220$ cm$^{-1}$) and $\omega=\sqrt{2}\times 10^{-3} E_h/\hbar$ for the restraining and for the NN oscillators, respectively, thus yielding identical force constants $m\nu^2=\mu\omega^2=10^{-6}E_h/a_0^2$.
In all our simulations, we employ a basis set consisting of $d^{(\mathrm{ex})}=2$ electronic and $d^{(\mathrm{ph})}=8$ vibrational basis functions.
The latter value has been obtained by testing the convergence with respect to this cut-off carefully.
Moreover, we consider here the symmetric exciton-phonon coupling mechanism of Eq.~\eqref{eq:H_couple_sig} which is typically used in modeling OSC materials.
The corresponding coupling constant, $\sigma$, is chosen one order of magnitude smaller than the vibrational frequencies, i.e., a few times $10^{-4} E_h/a_0$, for details see below.
Unless stated otherwise, we use a shift value of $\eta=0$ for the ALS eigensolver, i.e., we start searching eigenvalues at zero energy.

\subsection{Software}
\label{sec:results_soft}

All simulations presented in the following are carried out using the newly developed Python package \textsc{ExciDyn} which is available from one of the authors (B.~S.) upon request and which will be made publicly available in the near future. 
It encompasses various numerical solvers for the TISE and TDSE for Hamiltonians in TT form using a SLIM representation. 
For the underlying computations with tensors in the TT format, it uses solvers from \textsc{Scikit-TT}~\cite{Gelss2021}, an open-source tensor train toolbox for Python based on NumPy and SciPy.
Analysis and visualization of eigenstates in Sec.~\ref{sec:results_trap} is based on reduced densities.
An efficient implementation of the necessary tracing over all but one degrees of freedom is also available within \textsc{ExciDyn}.
From the reduced densities, $\rho_i$ it is straight-forward to obtain expectation values of observables of interest, most importantly averaged quantum numbers, $\langle n_i^\mathrm{(ex)}\rangle = \mathrm{tr}(\rho_i,b_i^\dagger b_i)$ and $\langle n_i^\mathrm{(ph)} \rangle = \mathrm{tr}(\rho_i,c_i^\dagger c_i)$, characterizing the excitonic and phononic states of the $i$-th site, respectively. 
The total degrees of excitation are characterized by summing over all sites, 
$\mathcal{N}^\mathrm{(ex)} \equiv  \sum_{i=1}^N \langle n_i^\mathrm{(ex)}\rangle$
and $\mathcal{N}^\mathrm{(ph)} \equiv \sum_{i=1}^N \langle n_i^\mathrm{(ph)}\rangle$.

For small numbers of sites, $N\le 5$, our results, obtained with \textsc{ExciDyn}, are in excellent agreement with values obtained from conventional, grid-based solvers for general Hamiltonians available within the \textsc{WavePacket} software package \cite{Schmidt2017,Schmidt2018,Schmidt2019}.

\subsection{Purely excitonic chains}
\label{sec:results_ex}
 
For the case of homogenous excitonic systems with $\alpha_i=\alpha$ and $\beta_i=\beta$ for all $1 \le i \le N$, the time-independent Schr{\"o}dinger equation (TISE) for the excitonic Hamiltonian \eqref{eq:H_ex} can be solved analytically, in close analogy to H{\"u}ckel theory.
Within the Fock space of singly excited states, i.e., with exciton number $\mathcal{N}^\mathrm{(ex)} = 1$, the eigenenergies are given by
\begin{equation}
  E_k^\mathrm{(ex)} = \alpha + 2 \beta \cos \left( ka \right)
	\label{eq:E_ex}
\end{equation}
where $a$ is the lattice constant.
The corresponding (Bloch type) eigenstates are the simplest case of a Bethe \textit{ansatz} 
\begin{equation}
  |\psi_k^\mathrm{(ex)}\rangle = N^{-1/2} \sum_{s=1}^N e^{ikas} b_s^\dagger |0\rangle
	\label{eq:psi_exact} 
\end{equation}
where $|0\rangle$ stands for the excitonic ground state.
For periodic systems, the wavenumbers $k_j=2\pi j/(aN)$ are discrete with $-\frac{N}{2}+1\leq j \leq \frac{N}{2}$ (for $N$ even) or $-\frac{N-1}{2}\leq j \leq \frac{N-1}{2}$ (for $N$ odd).
For non-periodic systems, one has to choose $k_j=\pi j / (a(N+1))$ with $1 \leq j \leq N$ instead.
Solutions for higher excitation numbers $\mathcal{N}^\mathrm{(ex)}$ can also be found in the recent literature but shall be not considered here \cite{Hu2018}. 

\begin{figure}[htbp]
\centering
\includegraphics[width=1\textwidth]{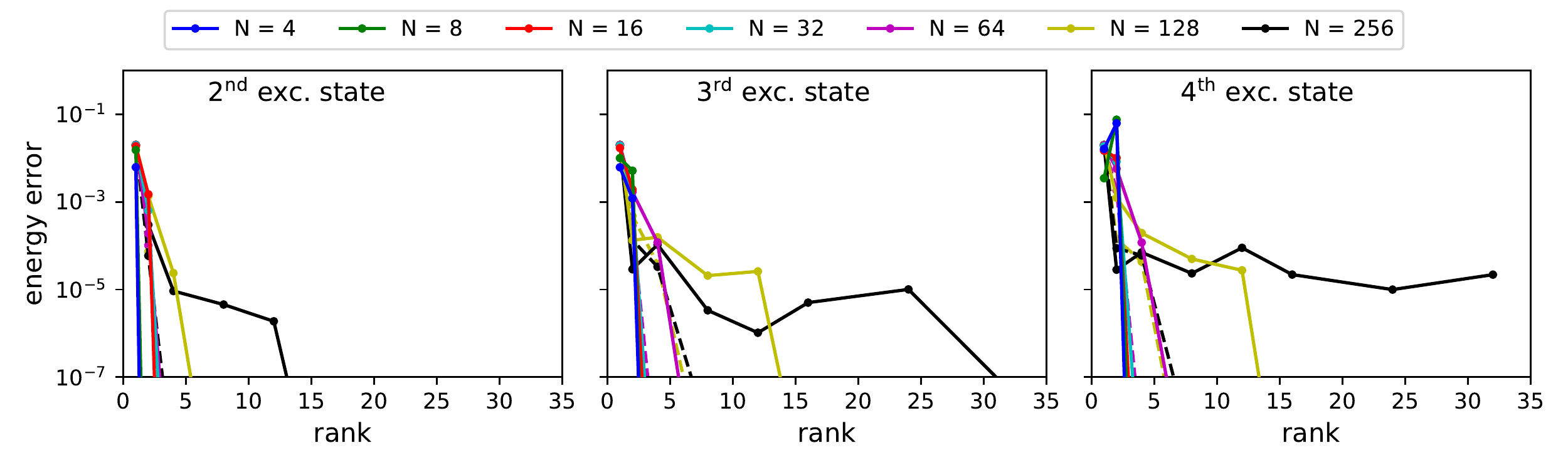}
\caption{Convergence with the rank of the error in the energy of the second to fourth excited states for a chain of excitons for different chain lengths $N$. Left: second excited state, middle: third excited state, right: fouth excited state. Solid lines: $s=256$, dashed lines: $s=8192$ (maximum number of ALS sweeps).}
\label{fig:ConvergenceExcitons}
\end{figure}

To test our TT-based numerical approach, we consider homogeneous, non-periodic excitonic chains of lengths $4\leq N \leq 256$, with the standard parameters given in Sec.~\ref{sec:results_gen}. 
For all chain lengths, we approximated the four lowest excited states, which have been obtained from the integrated Wielandt deflation described in Sec.~\ref{subsec:IntegratedWielandt}. 
For the values of $N$ considered here, this is the maximum number of excited states for which all states are from the Fock space of singly excited states.
We therefore expect that these allow for a fair comparison between different chain lengths. 
To address the convergence with respect to the maximum rank of the solution, we employed different values for $r \in \{1,2,4,8,12,16,24,32\}$, where, for one choice of $r$ and $N$, all five eigenstates are approximated by a TT of that rank. 

In all simulations, we terminated the ALS when the estimated eigenvalue did not change by more than $10^{-10}$ in the last three ALS sweeps, or when the number $s$ of ALS sweeps exceeded $256$. 
We observed that the ALS estimates of the eigenvalues are not always strictly decreasing after every new sweep.
We therefore employ the lowest value from the iteration and the corresponding TT as estimate for the eigenvalue and eigenstate, because this is the best approximation according to the Rayleigh quotient.
Note that the extra computational costs due to the modified Wielandt deflation were so moderate that the observed variations of the CPU-times on the employed compute cluster had a more significant impact. 
In all cases, the CPU-times stayed below $20$ hrs/state on a single thread of a {\sc Xeon Skylake 6130} CPU and most cases came at a small fraction of that cost (cf.~Sec.~\ref{subsec:IntegratedWielandt}).
This must be contrasted to the simple approach to Wielandt deflation, where we conduct the addition in Eq.~\eqref{eq:deflated_H} directly on the Hamiltonian before entering the ALS.
The concomitant increase in the rank then led to significant (often prohibitive) increases of the computational costs for the excited states already at modest chain length and ranks.

Fig.~\ref{fig:ConvergenceExcitons} summarizes our findings with respect to the influence of $N$ and $r$  and displays the convergence of the error of the energy of the second to fourth excited states with respect to the employed maximum rank $r$. 
We omit the discussion of the ground and the first excited state, because the former one is exact in the Hartree ($r=1$) approximation and the latter one is approximated to machine precision already at $r=2$.
While we find significant deviations for ranks $r=1$, increasing the rank leads to a fast decay of the error. 
Notably, if we set the maximum number of ALS sweeps to $s=256$, chains with $N\leq 32$ display negligible errors for ranks $r \leq 4$ for all considered excited states and the required rank to achieve a certain target accuracy seems to be more or less independent of the system size.
For longer chains with $N \geq 64$ (and $s=16$), this rule seems not to hold anymore and the decrease of the error is significantly slower with increasing $N$.
This becomes more pronounced for higher excited states, such that the case $N=64$ still behaves like the shorter chains for the second excited state while deviating at higher excited states. 
For the longest chain with $N=256$, we still see convergence for the second excited state, which slows down for the third excited state, and, for the fourth excited state, we observe no convergence anymore if we do not allow more than $256$ ALS sweeps. 
That is, for all cases which significantly deviate from the common fast convergence behavior, ALS never reaches our convergence criterion before it terminates. 

Indeed, setting the maximum number of sweeps to $s=8192$, we find that all cases show energy errors below the displayed range already for rank $r=8$ and we arrive at the desired, (almost) size independent convergence behavior (see the dashed lines in Fig.~\ref{fig:ConvergenceExcitons}).
Hence, the disturbed convergence with respect to the rank is an effect of an insufficiently converged ALS iteration.
The increased difficulty to converge the ALS for longer chains is not unexpected. 
With increasing chain length the gaps between the eigenvalues of the excited states decrease, which is known to negatively affect the convergence rate of iterative methods such as the Rayleigh quotient minimization. 
An explanation for the higher complexity of obtaining higher excited states might be that those involve higher wavenumbers, see Eq.~\eqref{eq:psi_exact}, and thus correlations are more long-ranged.
This is typically known to require higher TT ranks \cite{Gelss2016,Gelss2017}.

\subsection{Purely phononic chains}
\label{sec:results_ph}

As a second example, we again consider the case of homogenous systems, but now chains of oscillators with $m_i=m$, $\nu_i=\nu$, and $\omega_i=\omega$ for all $1 \le i \le N$.
For periodic systems, the TISE for the phononic Hamiltonian \eqref{eq:H_ph0} can be solved analytically.
By introducing (Bloch type) normal coordinates, $R_q=N^{-1/2}\sum_{s=1}^N e^{iqas}R_s$, and conjugated momenta, $P_q=N^{-1/2}\sum_{s=1}^N e^{-iqas}P_s$, with $q$ being the wavenumber of an acoustic phonon in one dimension, the Hamiltonian (\ref{eq:H_ph0}) can be rewritten as a sum of quasi-uncoupled normal mode harmonic oscillators
\begin{equation}
 	H^{\mathrm{(ph)}} = \frac{1}{2m}          \sum_{q=1}^N 
 	    \left(P_q P_{-q} + m^2 \Omega_q^2 R_q R_{-q} \right) \quad .						
 	\label{eq:H_ph1}
\end{equation}
The resulting eigenenergies are given by
\begin{equation}
E^{\mathrm{(ph)}}_{\mathbf n} = \sum_q \Omega_q \left( n_q + \frac{1}{2} \right) 
 	\label{eq:E_ph}
\end{equation}
where the $n_q\ge 0$ are integer phonon quantum numbers.
The corresponding frequencies $\Omega_q$ are given by 
\begin{equation}\Omega_q=\sqrt{\nu^2+\omega^2(1-\cos(qa))} 
\label{eq:omega_q}
\end{equation}
where the phonon wavenumbers $q$ take on the same values as the exciton wavenumbers $k$ given in Sec. \ref{sec:results_ex} for periodic chains. 

For non-periodic systems, however, fully analytic solutions are not available because of the non-uniformity of the effective frequencies of single sites ($\tilde{\nu}$) and NN pair ($\tilde{\omega}$) vibrations, see Eq.~\eqref{eq:omg_E}.
Instead, the frequencies in \eqref{eq:E_ph} are obtained from a conventional normal mode analysis as the square roots of the eigenvalues of the Hessian matrix of the phonon potential energy function in Eq.~\eqref{eq:H_ph0}.

\begin{figure}[htbp]
\centering
\includegraphics[width=1\textwidth]{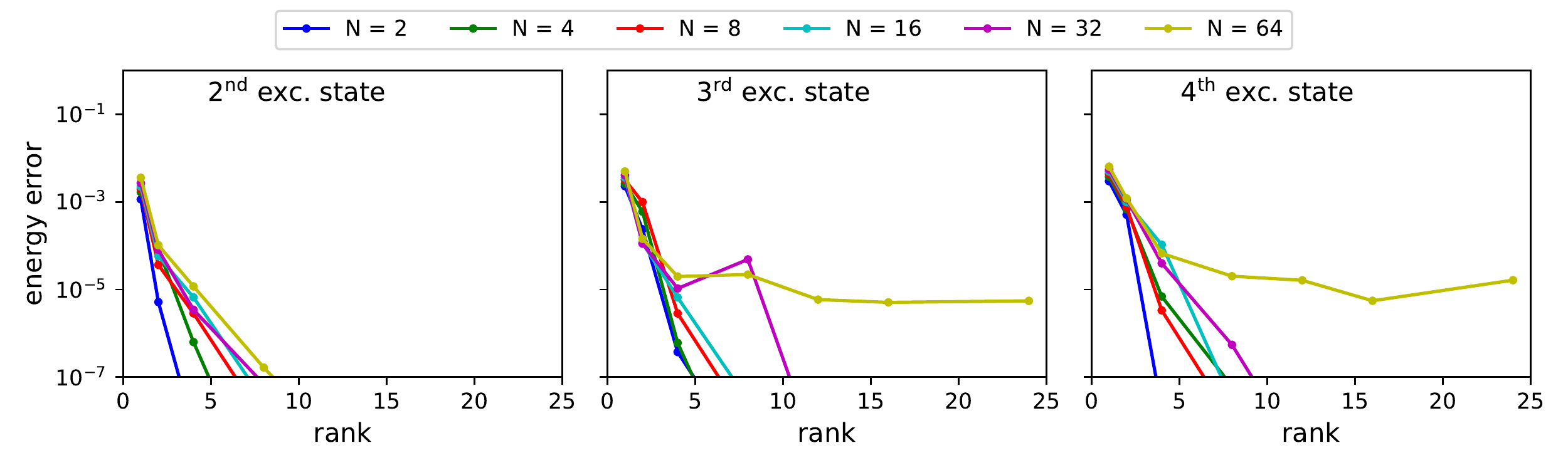}
\caption{Convergence with the rank of the error in the energy of the second to fourth excited states for a chain of oscillators for different chain lengths $N$. Left: second excited state, middle: third excited state, right: fouth excited state.}
\label{fig:ConvergencePhonons}
\end{figure}

For our numerical experiments, we considered non-periodic chains of lengths $4\leq N \leq 64$, employing the standard parameters from Sec.~\ref{sec:results_gen}. 
This choice for the maximum $N$ has the effect that the maximal dimensionality of the Hilbert spaces on which we are operating agrees with that of the excitonic case. 
All other settings are as in Sec.~\ref{sec:results_ex}, except that we have conducted simulations only for ranks $r \in \{1,2,4,8,12,16,24\}$ because of limitations of our computational resources. 
As before, we terminate after $s=256$ ALS sweeps unless we hit the convergence criterion of $10^{-10}$ before, and from all ALS iterations we select the best approximation to the Rayleigh minimization as estimates for the eigenpair. 
Again, the integrated Wielandt deflation added only negligible extra computational costs for the excited states compared to the ground state calculations.
In all cases, the computer time stayed below $35$ hrs/state on a single thread of the employed {\sc Xeon Skylake 6130} CPUs. 

Fig.~\ref{fig:ConvergencePhonons} displays the convergence of the energy error with the rank for the second (left), third (middle) and fourth eigenstates (right). 
As in Sec.~\ref{sec:results_ex}, we observe a very similar behavior. 
For increasing chain length, the convergence behavior seems to approach a common limiting behavior, such that the rank for a target accuracy is bound. 
Again, the observed deviations from this ideal behavior appear for higher excited states and longer chain length, where all these deviations come along with not reaching the convergence criterion and can therefore be expected to be caused by an insufficient number of ALS sweeps. 
Also in complete analogy to the previous section, the higher eigenvalues approach each other for increasing chain length, which explains the larger difficulties in converging the ALS. 
Due to higher vibrational wavenumbers involved, we expect excited states to be more delocalized which leads to stronger coupling between the sites and thereby more ALS sweeps to obtain a converged result.
Finally, notice that the ratio of site-to-site coupling versus on-site coupling for our model of the phonons ($\omega/\nu=\mathcal{O}(1)$) is much larger than for that of the excitons ($\beta/\alpha=0.1$).
Hence, in the former case the correlations are stronger which is the main reason why (moderately) higher ranks are required to achieve converged eigenvalues.

\subsection{Coupled phonons and excitons}
\label{sec:results_couple}

\begin{figure}[htbp]
\centering
\includegraphics[width=1\textwidth]{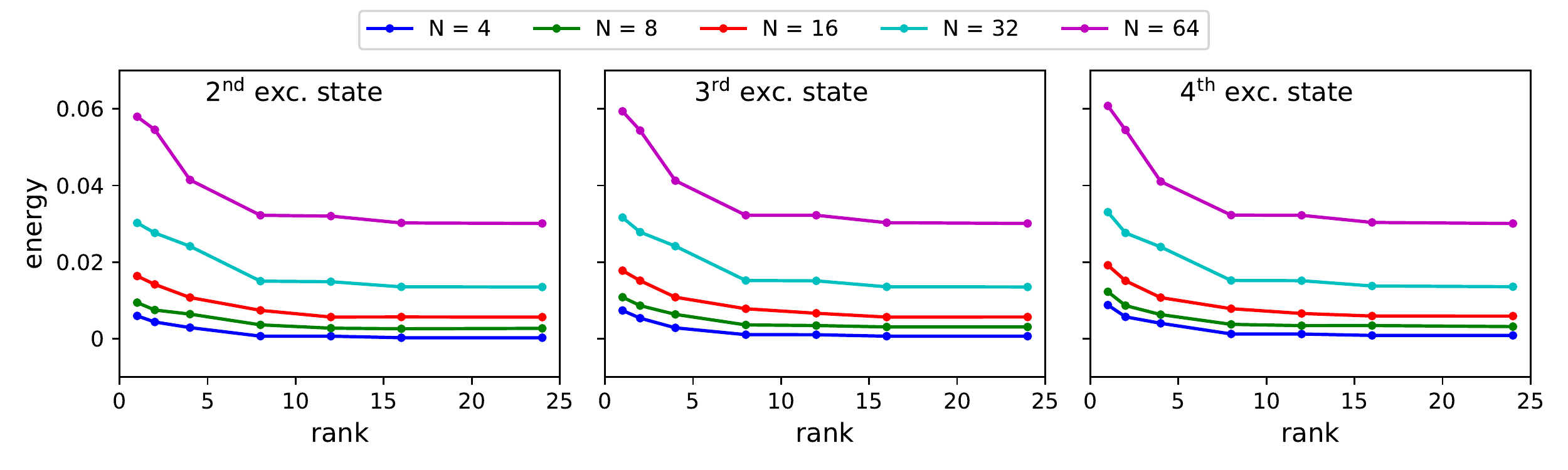}
\caption{Convergence of the eigenvalues of coupled exciton-phonon chains with the employed rank for varying chain lengths $N$. Left: second excited state, middle: third excited state,  right: fourth excited state.}
\label{fig:ConvergenceCoupled}
\end{figure}

The last convergence test is for a non-periodic, homogenous chain model of coupled phonons and excitons for chain lengths $4\leq N \leq 64$.
For the excitonic and phononic part of the Hamiltonian, we employ the default settings from Sec.~\ref{sec:results_gen}, except for the excitonic site energy, for which we choose a reduced value of $\alpha=0.021$.
The motivation for the alternate choice has been to artificially bring the lower band edge of the one-exciton states, $\alpha-2|\beta|$, just below the electronic ground state ($\mathcal{N}^{(ex)}$=0), thus partially separating it from the $\mathcal{N}^{(ex)}=1$ manifold.
Hence, we expect to find states with significant EPC effects already at low energies (choosing $\eta=0$) where they are amenable for testing our integrated Wielandt deflation approach.
Note that simulations with realistic values of $\alpha$ are presented in Sec.~\ref{sec:results_trap}.
For the symmetric EPC constant defined in Eq.~\eqref{eq:H_couple_sig} we choose $\sigma = 2 \times 10^{-4}$.

Fig.~\ref{fig:ConvergenceCoupled} displays the convergence of the energies of the second to fourth exited state. 
Note that, in absence of analytical solutions, we cannot display the energy error here but only the energies themselves on a linear instead of a logarithmic scale. 
While we still observe a rather fast convergence with the rank, we generally need higher ranks until the energy curves become completely flat compared to Sec.~\ref{sec:results_ex} and Sec.~\ref{sec:results_ph}, where, already at rank $r=4$, the errors are so low that they would be invisible on a linear scale. 
This might also explain why we do not observe significant differences between the three states. 
Moreover, while we observe a fast convergence in the energies, we do not see this for the residuals $||H \Psi_n - E_n \Psi_n||_2/N$.
We have no definite explanation, but the closeness of all states might indicate that the Hamiltonian is badly conditioned and the computation of the residuals might be affected by finite computational precision.

Besides these numerical results, we also inspected the physical nature of the converged states. 
For all cases, we find non-zero total quantum numbers $\mathcal{N}^\mathrm{(ex)}$ and $\mathcal{N}^\mathrm{(ph)}$ of excitons and phonons and these numbers become larger than one for chain lengths $N>4$.
Hence, the commonly employed single exciton approximations are not a good description for this problem anymore \cite{Schroter2015}.
The relatively strong EPC leads to the very small energies for low lying excited states for the coupled problem in comparison to the pure phonon problem.
Therefore, we envisage that in time-dependent scenarios quantum-classical approximations~\cite{Scott1992,Georgiev2019} will likely reach their limits and a full quantum description of all degrees of freedom, like the one presented here, is necessary.

Finally, we performed yet another test of our TT-based approach for a parameter set which is completely different from our exciton models for OSCs.
For the Holstein model of a polaron, i.e., an electron dressed by phonons, parameter values of $\beta=m=\nu=\chi=1$ and $\alpha=\omega=\rho=\sigma=\tau=0$ are suggested in Ref.~\cite{Bonca1999}.  
Among others, the authors of that study calculated eigenenergies for a chain length of $N=16$, employing exact diagonalization within a variational approach. 
We can reproduce their result for the stabilization energy up to five digits precision for a basis size of $d^{(\mathrm{ph})}=8$ and for $r=12$ within less than a minute on a mid-range PC.
\subsection{Example: Self-trapping effects}
\label{sec:results_trap}

Finally, we illustrate the potential of the TT--based methods for solving the TISE by presenting prototypical results on self-trapping (also known as self-localization or self-focusing) of excitons coupled to phonons, here longitudinal vibrations of a quasi one-dimensional chain. 
This effect is one of the hallmarks of the Fr\"{o}hlich-Holstein type Hamiltonians introduced in Sec.~\ref{sec:model}~\cite{Devreese2009}.
It was first predicted for the case of polarons by Landau as early as 1933, later also for soliton-like localization of coupled CO-stretching (amide I) vibrations along protein $\alpha$-helices \cite {Davydov1985,Scott1991,Scott1992}.
In particular, we focus here on the analytic solutions by A.~S.~Davydov, originally developed for the latter class of systems.
Assuming the quantum state of the coupled system to be a direct product of excitonic and phononic states, coupled equations of motion for excitons and phonons were derived within a mean-field (or Ehrenfest or time-dependent Hartree) formalism \cite{Kerr1987,Georgiev2019}.
There, the quantum state of the excitonic subsystem depends on the vibrational state via the EPC part of the Hamiltonian.
In turn, the quantum state of the phonon displacements is subject to a potential obtained as a mean-field from the state of the excitons. 
Assuming the vibrational degrees of freedom to relax to their corresponding equilibrium positions, Davydov obtained self-trapped or soliton solutions from a discretized version of the cubic non-linear Schr\"{o}dinger equation (NLSE)
\begin{equation}
E_b a_i = -\frac{|\beta|}{w}\left| a_i\right|^2 a_i - |\beta| \left( a_{i-1}-2a_i+a_{i+1}\right)
\label{eq:NLSE}
\end{equation}
for the expansion coefficients, $a_i$, of the excitonic subsystem.
The dimensionless width parameter $w$ is found to be inversely proportional to the square of $\sigma$, the parameter of the symmetric EPC introduced in Eq.~\eqref{eq:H_couple_sig}. 
Employing a continuum limit approximation for the NLSE for $N\gg 1$ and $w\gg 1$, Eq.~\eqref{eq:NLSE} can be solved analytically, and the ground state is given by 
\begin{equation}
a_i \propto \mathrm{sech} \left( \frac{i-i_0}{4w} \right) 
\label{eq:sech}
\end{equation} 
which describes a bell-shaped stationary wave packet centered around site $i_0$ with a full width at half maximum FWHM $\approx 7.05w$ in units of the lattice constant \cite{Scott1992}.
The corresponding soliton binding energy $E_b$ is proportional to the fourth power of the EPC constant $\sigma$.
The total soliton energy $E$ is obtained from the binding energy by subtracting the energy associated with the lattice deformation which displays the same type of power law.  
Finally, the overall stabilization $\Delta E\equiv E(\sigma)-E(0)$ is the difference of the total soliton energy and the energy of the uncoupled state.
The latter one is obtained as a sum of the lower band edge of the excitons from Eq.~\eqref{eq:psi_exact} with $k_m=0$ for $\beta<0$ (or $k_m=\pi/a$ for $\beta>0$) and the phononic zero point energy,
\begin{equation}
E(0)\equiv E_{k_m}^\mathrm{(ex)} + E_\mathbf{0}^\mathrm{(ph)} =\alpha-2|\beta|+\frac{1}{2}\sum_q \Omega_q \quad,
\label{eq:E0}
\end{equation}
see Eqs.~\eqref{eq:E_ex} and \eqref{eq:E_ph}.

In order to assess the validity of the Davydov theory and its underlying approximations, we conducted a series of simulations for periodic, homogenous systems with the number of sites between $N=15$ and $N=60$ and for the exciton and phonon energy parameters specified in Sec.~\ref{sec:results_gen}. 
Based on the convergence tests presented above, we set the maximal rank of the solutions, as well as the maximum number of ALS iterations to 20. 
Aiming at stabilization by self-trapping, we search eigenstates with electronic excitation $\mathcal{N}^\mathrm{(ex)} = 1$ and phononic excitation $\mathcal{N}^\mathrm{(ph)}$ as low as possible.
In the practice of our simulations, however, we also encounter states with $\mathcal{N}^\mathrm{(ex)} = 0$ and rather high phononic excitation $\mathcal{N}^\mathrm{(ph)}$ which coexist with the wanted states at energies just below $E(0)$ and which are discarded manually.  
This problem can be mitigated once a good estimate for $E(\sigma)$ is available and which is then used as energy shift parameter $\eta$ for the ALS eigensolver.

\begin{figure}[htbp]
\centering
\includegraphics[width=0.6\textwidth]{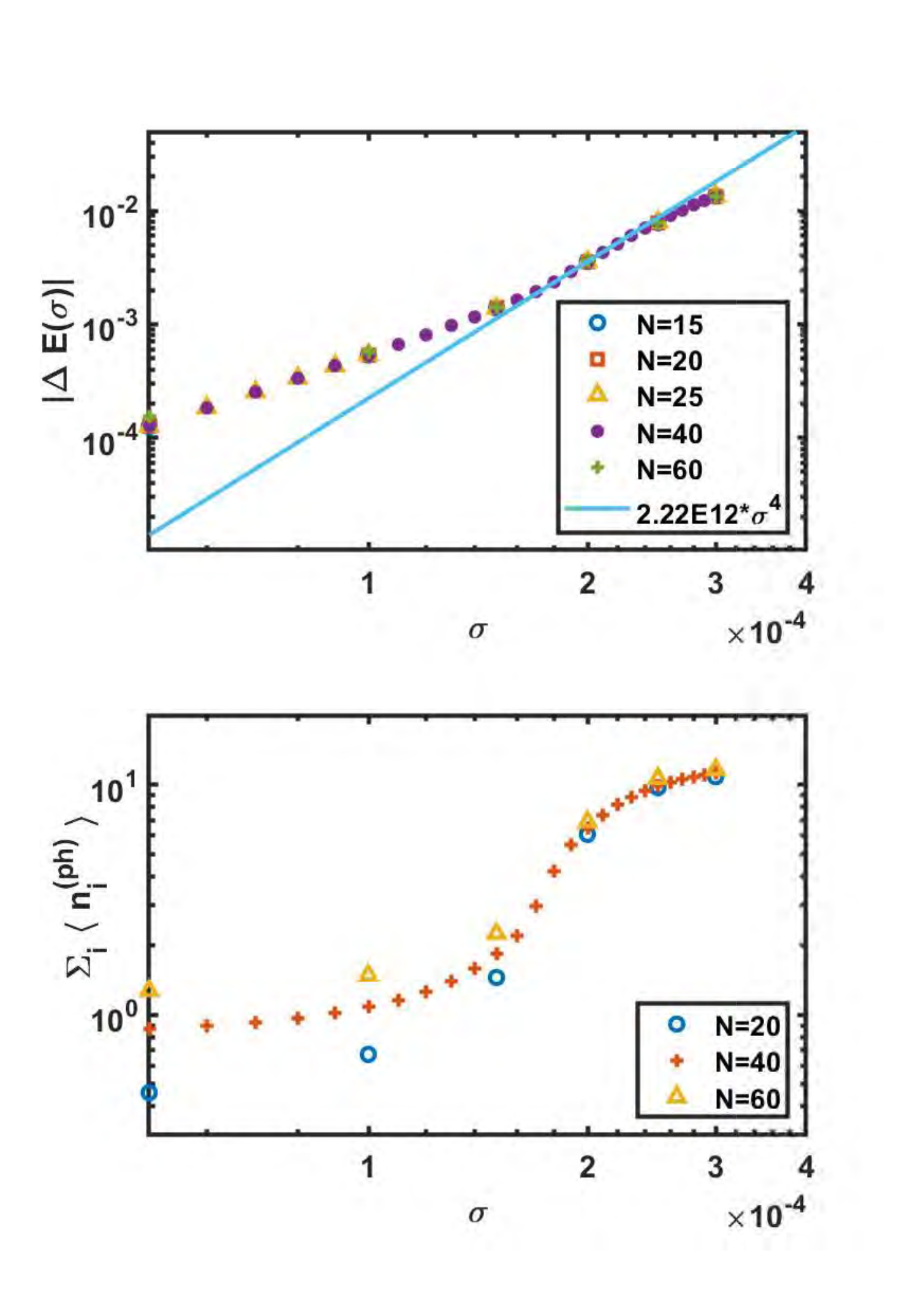}
\caption{Stabilization by self-trapping of excitons and phonons due to symmetrized EPC constant $\sigma$. 
Data points show results obtained with the TT method as described in this work, for various values of the chain length $N$. 
Top: Energetic stabilization $\Delta E=E(\sigma)-E(0)$, with the straight line representing a fourth power law fitted to the data around $\sigma=2\times10^{-4}$. 
Bottom: Total degree of excitation of lattice vibrations.}
\label{fig:Esig}
\end{figure}

\begin{figure}[htbp]
\centering
\includegraphics[width=0.6\textwidth]{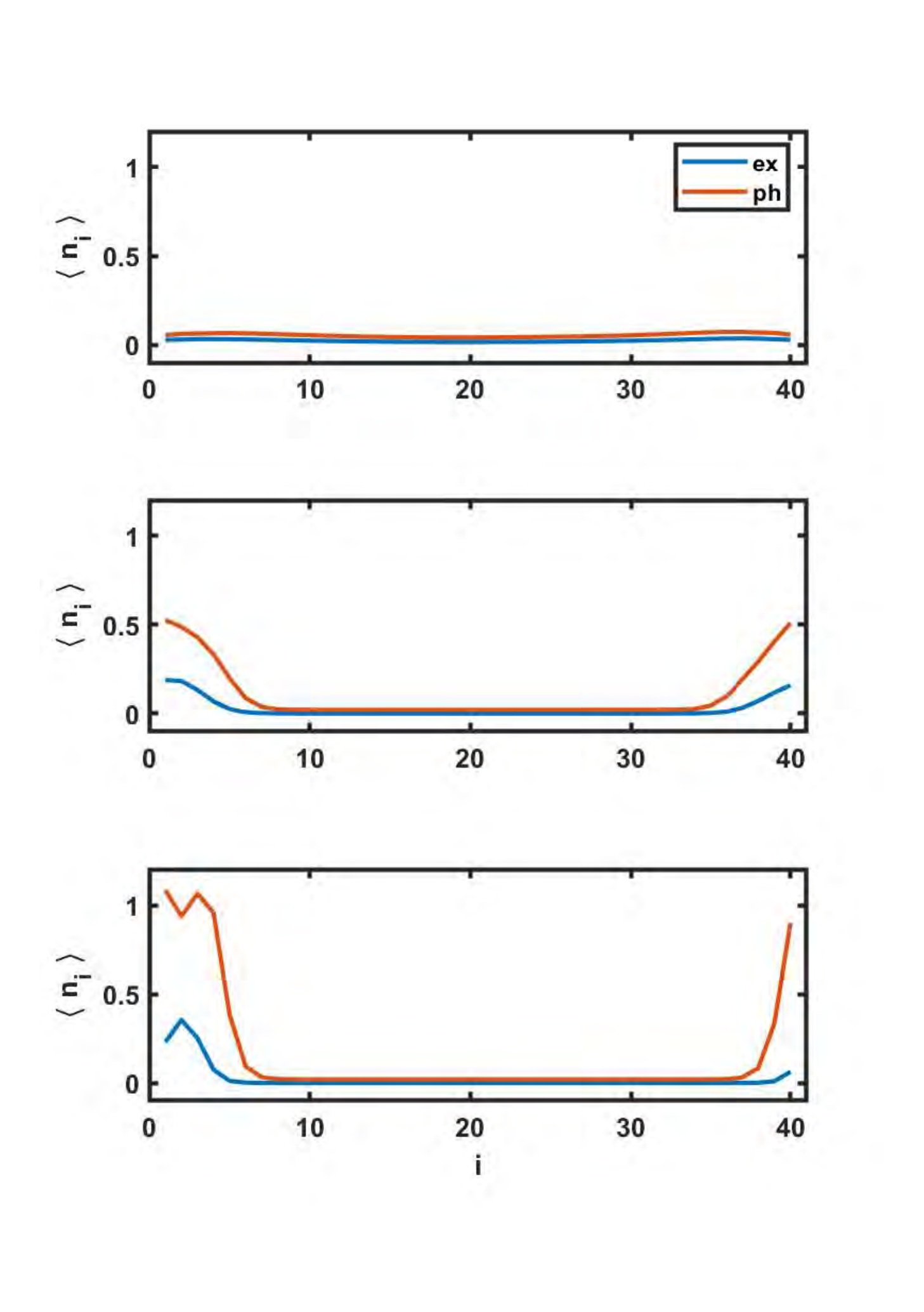}
\caption{Onset of self-localization of excitons and phonons for a chain length of $N=40$: Quantum numbers of excitons and phonons for each site and for various values of the symmetrized EPC constant.
Top: $\sigma=1.6 \times 10^{-4}$.
Middle: $\sigma=1.8 \times 10^{-4}$.
Bottom: $\sigma=2.0 \times 10^{-4}$.}
\label{fig:N40sig}
\end{figure}

\begin{figure}[htbp]
\centering
\includegraphics[width=0.6\textwidth]{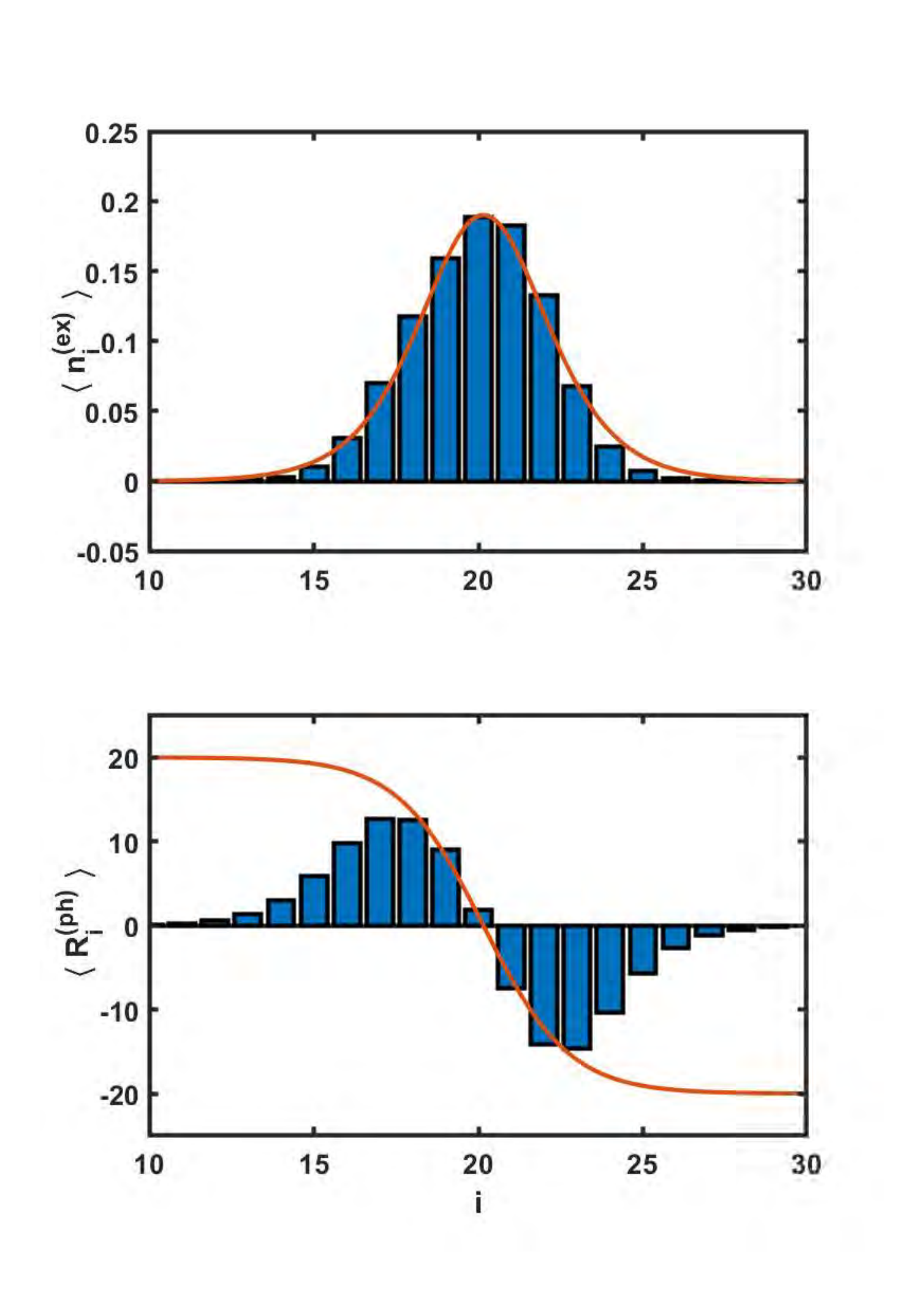}
\caption{Details of self-localization of electrons and phonons for a chain length of $N=40$ and for EPC constant $\sigma=1.8 \times 10^{-4}$.
Top: Quantum numbers of excitons (blue bars), together with a sech$^2$ fit function (red curve).
Bottom: Local distortions of the lattice sites (blue bars), together with a $1-$tanh$^2$ fit function (red curve).}
\label{fig:N40sech}
\end{figure}

The dependence of the stabilization energy $\Delta E$ on the EPC parameter $\sigma$ is shown in the upper panel of Fig.~\ref{fig:Esig}.
Within the range $1.7 \times 10^{-4} \le \sigma \le 2.3 \times 10^{-4}$, the values of $\Delta E$ are indeed very close to a fourth power law as suggested by the Davydov theory.
Next, we consider Fig.~\ref{fig:N40sig} representing the expected quantum numbers of excitons, $\langle n_i^{\mathrm{(ex)}}\rangle$, and phonons, $\langle n_i^{\mathrm{(ph)}}\rangle$ for each of the sites $i$.
These data confirm that the lowering of the energy indeed coincides with the mutual self-trapping of excitons and phonons in the same narrow region of the chain.
Moreover, Fig.~\ref{fig:N40sech} (upper panel) confirms that the probability distribution of the former ones closely follows a sech$^2$ behavior as predicted by the Davydov theory. 
In contrast, for $\sigma>2.3 \times 10^{-4}$ the stabilization energy (upper panel of Fig.~\ref{fig:Esig}) stays below the fourth power law, obviously because one of the basic assumptions for the continuum limit approximation ($w\gg 1$) is violated, i.e., the predicted width of the wave packet is not much larger than the lattice constant any more. 
Also for $\sigma<1.7 \times 10^{-4}$ there are notable deviations from the predictions of the Davydov theory. 
One might assume that this is because the predicted wave packets become too wide to be accommodated on a finite chain length $N$.
However, this is not the case because our simulations show that the deviations appear to be independent of $N$, see again the upper panel of Fig.~\ref{fig:Esig}.
Instead, there is a sudden onset of self-trapping around $\sigma=1.8\times 10^{-4}$, whereas for $\sigma \le 1.6\times 10^{-4}$, excitonic and vibrational occupation numbers is still practically equal for all sites, see Fig.~\ref{fig:N40sig}.
Also note that the sudden onset of self-localization is accompanied by a sudden rise of the total quantum number of lattice vibrations, $\mathcal{N}^\mathrm{(ph)}$, as shown in the lower panel of Fig.~\ref{fig:Esig}, whereas the total quantum number of excitons, $\mathcal{N}^\mathrm{(ex)}$, remains very close to unity for all values of $\sigma$ considered here (not shown in the figures).
The reason for these deviations of our TT-based results from the predictions of the Davydov theory is believed to be due to the additional restraining $\nu$ oscillators in our model, see Eq.~\eqref{eq:H_ph0} which prevent self-trapping for weak EPC cases.
Another consequence of these restraints is shown in the lower panel of Fig.~\ref{fig:N40sech}. The lattice distortions caused by the coupling to the excitons do not follow a $1-$tanh$^2$ behavior as predicted by the Davydov theory but rather tend to zero for sites far enough from center of the soliton. 

\section{Conclusions and Outlook}
\label{sec:conclude}
In this paper, we have shown how to apply the tensor-train format to solve the time-independent Schr\"{o}dinger equation for coupled excitons and phonons. 
The aim of the present approach is to mitigate the curse of dimensionality, i.e., to reduce the memory consumption as well as the computational costs significantly, at least for problem classes with a suitable mathematical structure.
Based on Fr\"{o}hlich-Holstein type model Hamiltonians, we have employed SLIM decompositions to construct TT representations of the corresponding Schr\"{o}dinger equations. 
Using this kind of systematic decomposition, we can easily scale the system size and the parameters without manually constructing the Hamiltonians in TT format for each setting.
Another advantage is that in case of homogeneous systems the TT ranks of the corresponding operators do not depend on the network size, which results in a linear growth of the storage consumption.
In order to compute quantum states and corresponding energies given by eigenpairs of the tensorized Hamiltonian, we combined ALS with the Wielandt deflation technique. 
While this increases the computational costs only slightly compared to ALS for approximating the ground state, we are able to compute a set of quantum states corresponding to the lowest energies (above a certain level).

We have tested the approach on three different classes of homogeneous chains for which we conducted an extensive investigation of the influence of the TT ranks of the approximate eigenstates and the chain length. 
Theses classes are purely excitonic chains, purely phononic chains and chains of excitons and phonons with a strong EPC, all non-periodic and homogeneous. 
Our results indicate that the rank to achieve a certain target accuracy is largely independent of the chain length. 
Together with the bounded operator rank guaranteed by the SLIM decomposition, this has resulted in a linear complexity in the chain length in terms of computational storage. 
Employing the same number of ALS sweeps in all cases, this translates directly into a a linear complexity in terms of CPU time. 
However, we found that longer chains require more sweeps to obtain converged results. 
As the CPU time per state is linear in the number of sweeps, this means that the optimal strategy shows a slight super-linear complexity. 
Although the approach already showed its efficiency, there is plenty of room for improvement. 
First of all, this concerns the employed eigenproblem solvers for the sub-problems in the ALS, for which we simply employed off-the-shelve SciPy routines which leads to an unfavorable scaling with respect to the ranks  and the dimensionality $d$ per site, cf.~Sec.~\ref{sec: ALS}. 
Using (dedicated) iterative solvers and exploiting the sparsity of the TT operator cores, this might be substantially improved. 
If the dimension $d$ gets large, the computational burden can further be reduced by employing the so-called quantized tensor-train (QTT) decompositions~\cite{Khoromskij2011}. 
That is, the TT cores of the Hamiltonians as well as of the quantum states are divided into smaller cores. 
If the resulting coupling ranks are sufficiently small, the use of the QTT format can be advantageous in terms of numerical efficiency and, thus, may enable us to mitigate the curse of dimensionality for even higher dimensions.
The QTT format has already been implemented into \textsc{Scikit-TT} and \textsc{ExciDyn}.
However, the ALS scheme is intrinsically sequential and therefore a straight-forward parallelization is only possible on the level of the sub-problems which are solved for every core. 
If these sub-problems become large (large $d$, large $r$), this might still be worth the effort.
Indeed, for the coupled cases, we have utilized the intrinsic multi-threading capabilities of NumPy and SciPy, which led to substantial reductions of the run time. 
Altogether,  we expect that our approach can be extended to way more complex scenarios than the ones investigated in this study. 
However, we do not want to conceal the limitations of our approach. 
By construction, Tensor Trains represent a chain of sites and, in our experience, they are substantially less efficient for problems where the operators do not encode a chain-like interaction.
In such cases, resorting to more flexible tensor network topologies such as the Hierarchical Tucker format  -- known in the context of quantum dynamics as \emph{multi-layer multi-configuration time-dependent Hartree} (ML-MCTDH)~\cite{Beck2000,Wang2003,Meyer2009} -- will likely help, but comes typically at the cost of more complex data structures and many adjustable parameters.

Finally, we have demonstrated that the TT-based quantum treatment of excitons and phonons makes it possible to investigate the phenomenon of mutual self-trapping without having to resort to the frequently used quantum-classical approximation~\cite{Georgiev2019} and without the rather restrictive assumption of a product structure for the state vectors of the coupled system.
Thus, we have been able to confirm the main results of the Davydov theory, though only for a certain range of the EPC parameter $\sigma$.
However, it is emphasized that our numerical techniques allow to perform calculations also beyond the validity regime of that theory, e.g., toward shorter chains and/or toward weaker or stronger EPC interactions.

It is also worth mentioning that the modeling assumptions underlying the Fr\"{o}hlich-Holstein type Hamiltonians introduced in Sec.~\ref{sec:model} are rarely fully justified in OSCs. 
In other words, Eqs.~(\ref{eq:H_ex}--\ref{eq:H_couple_tau}) are only a minimal model of exciton dynamics in real world materials. 
However, the approach of the present work could be used to overcome the basic limitations of the Fr\"{o}hlich-Holstein type modeling, such as the harmonic approximation for the phonon modes and the linear approximation for the EPC function.
Extensions such as anharmonic vibrational force fields or non-linear EPC functions can be treated within our TT-based formalism as long as they can be expressed in terms of the SLIM decomposition \eqref{eq:SLIM_1} for a tensor train with elementary tensors for on-site ($S$) and NN ($L,M$) contributions only.
Note that the underlying Hamiltonians do not necessarily have to be expressed in second quantization as presented here; other discretization schemes such as pseudo-spectral representations are also admissible. 
Another obvious extension would be to go beyond NN interactions in order to include interactions with longer range \cite{Laskin2012}, which might be realized by considering groups of neighboring sites and constructing the corresponding supercores based on the interaction between them.
Further computational challenges arise when taking more than one vibrational degree of freedom per site into account.
This becomes mandatory, e.g., for a realistic description of the conformational dynamics of $\pi$-conjugated polymer chains~\cite{Binder2018,DiMaiolo2020}.
In the practice of our approach based on TT and ALS, this would require much larger values of $d^\mathrm{(ph)}$.
Finally, the simplified assumption of a single band gap $\alpha$ between valence and conduction bands, frequently found in computational studies of inorganic semiconductors, may be insufficient for modeling the complexity of electronic states in organic semiconductors.
Depending on the chemical OSC composition and the wavelength of the light used in the excitation process, not 
only Frenkel singlet but also triplet and/or charge transfer states may be playing a role~\cite{Bardeen2014}.
In the context of our TT approaches this would require larger values of $d^\mathrm{(ex)}$.
Both cases, i.e., increasing the dimension of the excitonic and phononic Hilbert spaces, may be approached by QTT decompositions, see above.

Apart from the question of the validity of the model equations to describe the photophysical/photochemical dynamics of certain materials, other challenging problems arise from the choice of the specific values for the parameters $\alpha, \beta, \nu, \omega, \chi, \rho, \sigma, \tau$ in our model Hamiltonians.
One choice is to consider them as empirical parameters, which are fitted to experimental data, typically from spectroscopy.
Alternatively, the parameters of Fr\"{o}hlich-Holstein type Hamiltonians for OSCs can be derived from first-principles quantum-chemical calculations ~\cite{Binder2018,DiMaiolo2020} or from quantum-mechanical/molecular mechanics~\cite{Kranz2016}.
Currently, efficient and reliable schemes to obtain electron-phonon coupling~\cite{Giustino2017} are  being transferred  exciton-phonon coupling~\cite{Chen2020}.

Finally, we mention that the current approach of treating EPC problems with a SLIM representation of TTs is about to be applied to  time-dependent problems as well. 
The development and implementation of fast and reliable solvers based on the ALS scheme for the TDSE will allow us to target the  quantum dynamics of excitonic energy transport in a forthcoming publication.

\begin{acknowledgments}
Funded by the Deutsche Forschungsgemeinschaft (DFG, German Research Foundation) under Germany's Excellence Strategy -- The Berlin Mathematics Research Center MATH+ (EXC-2046/1, project ID: 390685689) and by the CRC 1114 ``Scaling Cascades in Complex Systems'' funded by the Deutsche Forschungsgemeinschaft (project ID: 235221301).
Felix Henneke (FU Berlin) and Reinhold Schneider (TU Berlin) are acknowledged for insightful discussions and Jerome Riedel for valuable help with the implementation of the Python codes. 
Moreover, the authors would like to thank the HPC Service of ZEDAT, FU Berlin, for generous allocation of computing resources.
\end{acknowledgments}

\section*{Data Availability}
The data that support the findings of this study are available from the corresponding author upon reasonable request.

\newpage

\bibliography{TISE_TT_1} 

\appendix

\section{Construction and normalization of TT cores}
\subsection{Orthonormalization}
\label{app: orthonormalization}

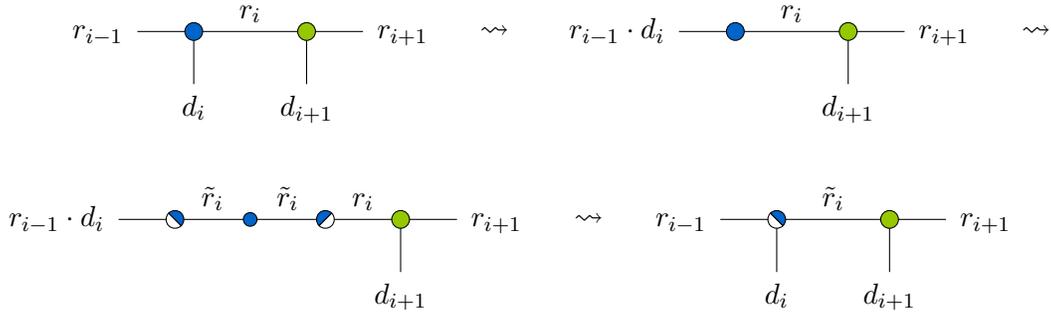
\begin{figure}[htbp]
\centering

\begin{tikzpicture}
\draw[black] (-0.75,0) -- node [label={[shift={(-0.9,-0.45)}]$r_{i-1}$}] {} ++ (0.75,0) ;
\draw[black] (0,0) -- node [label={[shift={(0,-0.15)}]$r_i$}] {} ++ (1.5,0) ;
\draw[black] (1.5,0) -- node [label={[shift={(0.9,-0.45)}]$r_{i+1}$}] {} ++ (0.75,0) ;
\draw[black] (0,0) -- node [label={[shift={(0,-1.1)}]$d_i$}] {} ++ (0,-0.7) ;
\draw[black] (1.5,0) -- node [label={[shift={(0,-1.13)}]$d_{i+1}$}] {} ++ (0,-0.7) ;
\node[draw,shape=circle,fill=Blue, scale=0.65] at (0,0){};
\node[draw,shape=circle,fill=Green, scale=0.65] at (1.5,0){};
\node[] at (4,0) {$\rightsquigarrow$};
\draw[black] (6.45,0) -- node [label={[shift={(-1.2,-0.45)}]$r_{i-1} \cdot d_i$}] {} ++ (0.75,0) ;
\draw[black] (7.2,0) -- node [label={[shift={(0,-0.15)}]$r_i$}] {} ++ (1.5,0) ;
\draw[black] (8.7,0) -- node [label={[shift={(0.9,-0.45)}]$r_{i+1}$}] {} ++ (0.75,0) ;
\draw[black] (8.7,0) -- node [label={[shift={(0,-1.13)}]$d_{i+1}$}] {} ++ (0,-0.7) ;
\node[draw,shape=circle,fill=Blue, scale=0.65] at (7.2,0){};
\node[draw,shape=circle,fill=Green, scale=0.65] at (8.7,0){};
\node[] at (11.2,0) {$\rightsquigarrow$};
\def\x{-1}
\draw[black] (\x,-2.5) -- node [label={[shift={(-1.2,-0.45)}]$r_{i-1} \cdot d_i$}] {} ++ (0.75,0) ;
\draw[black] (\x+0.75,-2.5) -- node [label={[shift={(0,-0.15)}]$\tilde{r}_i$}] {} ++ (1,0) ;
\draw[black] (\x+1.75,-2.5) -- node [label={[shift={(0,-0.15)}]$\tilde{r}_i$}] {} ++ (1,0) ;
\draw[black] (\x+2.75,-2.5) -- node [label={[shift={(0,-0.15)}]$r_i$}] {} ++ (1,0) ;
\node[draw,shape=semicircle,rotate=135,fill=white, anchor=south,inner sep=2pt, outer sep=0pt, scale=0.75] at (\x+0.75,-2.5){}; 
\node[draw,shape=semicircle,rotate=315,fill=Blue, anchor=south,inner sep=2pt, outer sep=0pt, scale=0.75] at (\x+0.75,-2.5){};
\node[draw,shape=circle,fill=Blue, scale=0.5] at (\x+1.75,-2.5){};
\node[draw,shape=semicircle,rotate=45,fill=Blue, anchor=south,inner sep=2pt, outer sep=0pt, scale=0.75] at (\x+2.75,-2.5){}; 
\node[draw,shape=semicircle,rotate=225,fill=white, anchor=south,inner sep=2pt, outer sep=0pt, scale=0.75] at (\x+2.75,-2.5){};
\draw[black] (\x+3.75,-2.5) -- node [label={[shift={(0.9,-0.45)}]$r_{i+1}$}] {} ++ (0.75,0) ;
\draw[black] (\x+3.75,-2.5) -- node [label={[shift={(0,-1.13)}]$d_{i+1}$}] {} ++ (0,-0.7) ;
\node[draw,shape=circle,fill=Green, scale=0.65] at (\x+3.75,-2.5){};
\node[] at (\x+6.25,-2.5) {$\rightsquigarrow$};
\draw[black] (\x+8,-2.5) -- node [label={[shift={(-0.9,-0.45)}]$r_{i-1}$}] {} ++ (0.75,0) ;
\draw[black] (\x+8.75,-2.5) -- node [label={[shift={(0,-0.15)}]$\tilde{r}_i$}] {} ++ (1.5,0) ;
\draw[black] (\x+10.25,-2.5) -- node [label={[shift={(0.9,-0.45)}]$r_{i+1}$}] {} ++ (0.75,0) ;
\draw[black] (\x+8.75,-2.5) -- node [label={[shift={(0,-1.1)}]$d_i$}] {} ++ (0,-0.7) ;
\draw[black] (\x+10.25,-2.5) -- node [label={[shift={(0,-1.13)}]$d_{i+1}$}] {} ++ (0,-0.7) ;
\node[draw,shape=semicircle,rotate=135,fill=white, anchor=south,inner sep=2pt, outer sep=0pt, scale=0.75] at (\x+8.75,-2.5){}; 
\node[draw,shape=semicircle,rotate=315,fill=Blue, anchor=south,inner sep=2pt, outer sep=0pt, scale=0.75] at (\x+8.75,-2.5){};
\node[draw,shape=circle,fill=Green, scale=0.65] at (\x+10.25,-2.5){};
\end{tikzpicture}
\caption{Orthonormalization of TT cores: In order to left-orthonormalize a TT core $\Psi^{(i)} \in \mathbb{R}^{r_{i-1} \times d_i \times r_i}$ (blue circle), the core is first reshaped into a matrix in $\mathbb{R}^{(r_{i-1} \cdot d_i) \times r_i}$. 
Then, an SVD is applied to this matrix resulting in the matrices $U$, $\Sigma$, and $V^\top$ (half-filled and small blue circles, respectively). 
The component $U$ builds the updated version of $\Psi^{(i)}$ while $\Sigma$ and $V^\top$ are contracted with the core $\Psi^{(i+1)}$ (green circle). 
The new rank between $\Psi^{(i)}$ and $\Psi^{(i+1)}$ is $\tilde{r}_i$. 
Right-orthonormalization is analogous.}
\label{fig: ortho}
\end{figure}

Given a core of a tensor train, we reshape it into a matrix and then apply a (truncated) SVD. 
Depending on the direction of the orthonormalization, the left- or right-singular vectors then define the updated core and the non-orthonormal part is contracted with the next core, see Fig.~\ref{fig: ortho} where the left-orthonormalization of a TT core is shown. 

Note that we here use orthonormalization particularly to ensure stability of ALS, see Sec.~\ref{sec: ALS}, since ALS is by design not rank-adaptive in general. 
Without truncation, the procedure above would simply compute a different but equivalent TT representation. 
The TT ranks may decrease by left- or right-orthonormalization but they cannot increase since the updated TT ranks are bounded by the initial ranks. 
If we decompose a core as shown in Fig.~\ref{fig: ortho}, the resulting rank $\tilde{r}_i$ satisfies
\begin{equation}
 \tilde{r}_i \leq \min\{r_{i-1} \cdot d_i , r_i\}.
\end{equation}
Thus, assuming that all cores have the same mode dimension $d$, the maximum TT rank of a tensor train with $N$ cores (after left- and right-orthonormalization) is given by $d^{\left\lfloor N/2 \right\rfloor}$.

\subsection{SLIM supercores}
\label{app: SLIM supercores}

We gather all matrices $L_{i,\lambda}$ and $M_{i+1,\lambda}$ into TT cores $L_i$ and $M_{i+1}$, respectively. 
These cores are defined as
\begin{equation*}
    \begin{alignedat}{2}
        \llbracket L_i \rrbracket 
            & =
            \left\llbracket \begin{matrix} L_{i,1} &  \dots & L_{i,\xi_i} \end{matrix}\right\rrbracket
            && \in \mathbb{R}^{1 \times d_i \times d_i \times \xi_i }, \\
        \llbracket M_{i+1} \rrbracket 
            & =
            \left\llbracket\begin{matrix} M_{i+1,1}  & \dots & M_{i+1,\xi_i}  \end{matrix}\right\rrbracket^\mathbb{T}
            && \in \mathbb{R}^{\xi_i \times d_{i+1} \times d_{i+1} \times 1},
    \end{alignedat}
\end{equation*}
for $ i = 1, \dots, N-1 $, and
\begin{equation}
    \begin{alignedat}{2}
        \llbracket L_N \rrbracket 
            & =
            \left\llbracket\begin{matrix} L_{N,1}  & \dots & L_{N,\xi_N}  \end{matrix}\right\rrbracket^\mathbb{T}
            && \in \mathbb{R}^{ \xi_N \times d_N \times d_N \times 1}, \\
        \llbracket M_1 \rrbracket 
            & =
            \left\llbracket\begin{matrix} M_{1,1}  & \dots & M_{1,\xi_d}  \end{matrix}\right\rrbracket
            && \in \mathbb{R}^{1 \times d_1 \times d_1 \times \xi_N },
    \end{alignedat}
\end{equation}
for the case of cyclic systems only.
Utilizing the core-notation and the definition of rank-transposed TT cores from Ref.~\cite{Gelss2017}, the Hamiltonian tensor may be expressed as one compact tensor train. 
Thus, the whole operator can be written in a shorter form as
\begin{equation} 
    \begin{split}
        H &= S_1 \otimes I_2 \otimes \dots \otimes I_d  + \dots + I_1 \otimes \dots \otimes I_{d-1} \otimes S_d \\
        & \quad + \llbracket L_1  \rrbracket \otimes \llbracket M_2 \rrbracket \otimes I_3 \otimes \dots \otimes I_d + \dots + I_1 \otimes \dots \otimes I_{d-2} \otimes \llbracket L_{d-1} \rrbracket \otimes \llbracket M_d \rrbracket\\
        & \quad + \llbracket M_1 \rrbracket \otimes \llbracket J_2 \rrbracket \otimes \dots \otimes \llbracket J_{d-1} \rrbracket \otimes \llbracket L_{d} \rrbracket,
    \end{split}
		\label{eq:SLIM_2}
\end{equation}
where $J_i \in \mathbb{R}^{\xi_N \times d_i \times d_i \times \xi_N} $ is a TT core with
\begin{equation}
 \llbracket J_i \rrbracket = \left\llbracket \begin{matrix}
        I_i  &  & 0 \\
           & \ddots & \\
           0 & & I_i
    \end{matrix}\right\rrbracket.
\end{equation}
Note that here and below the square brackets do not stand for block matrices but for the compact representation of a tensor. 
As before, the last line of Eq.~\eqref{eq:SLIM_2} is only required for cyclic systems.

\end{document}